\documentclass[fleqn,usenatbib]{mnras}

\usepackage{newtxtext,newtxmath}

\usepackage[T1]{fontenc}

\DeclareRobustCommand{\VAN}[3]{#2}
\let\VANthebibliography\thebibliography
\def\thebibliography{\DeclareRobustCommand{\VAN}[3]{##3}\VANthebibliography}

\usepackage{graphicx}	
\usepackage{amsmath}	

\title[SMBH Accretion Rate Enhancements in Post-Mergers]{Interacting galaxies in the IllustrisTNG simulations - IV: Enhanced Supermassive Black Hole Accretion Rates in Post-Merger Galaxies}

\author[S. Byrne-Mamahit et al.]{
Shoshannah Byrne-Mamahit,$^{1}$\thanks{E-mail: sjbyrnem@uvic.ca}
Maan H. Hani,$^{2}$
Sara L. Ellison,$^{1}$
Salvatore Quai,$^{3,4}$
David R. Patton,$^{5}$
\\
$^{1}$ Department of Physics \& Astronomy, University of Victoria, 3800 Finnerty Road, Victoria, British Columbia, V8P 5C2, Canada\\
$^{2}$ Department of Physics \& Astronomy, McMaster University, 1280 Main Street W, Hamilton, Ontario, L8S 4K1, Canada\\
$^{3}$ Dipartimento di Fisica e Astronomia “Augusto Righi,” Universit\`{a} degli Studi di Bologna, Via Gobetti 93/2, I-40129 Bologna, Italy \\
$^{4}$ INAF-Osservatorio di Astrofisica e Scienze dello Spazio di Bologna, Via Gobetti 93/3, I-40129 Bologna, Italy \\
$^{5}$ Department of Physics \& Astronomy, Trent University, 1600 West Bank Drive, Peterborough, Ontario, K9L 0G2, Canada\\
}

\date{Accepted XXX. Received YYY; in original form ZZZ}

\pubyear{2022}

\begin{document}
\label{firstpage}
\pagerange{\pageref{firstpage}--\pageref{lastpage}}
\maketitle

\begin{abstract}
We present an analysis of the instantaneous supermassive black hole (SMBH) accretion rates in a collection of 1563 post-merger galaxies drawn from the IllustrisTNG simulation. Our sample consists of galaxies that have experienced a merger in the last simulation snapshot (within $\sim$ 160 Myrs of coalescence) in the redshift range $0<z<1$, with merger stellar mass ratios $>1:10$ and post-merger stellar masses $> 10^{10} \mathrm{M_{\odot}}$. We find that, on average, the accretion rates of the post-mergers are $\sim$1.7 times higher than in a control sample and that post-mergers are 3-4 times more likely to experience a luminous active galactic nuclei (AGN) phase than isolated galaxies.  SMBH accretion rate enhancements persist for $\sim 2$ Gyrs after coalescence, significantly exceeding the $\sim$500 Myr lifetime of star formation rate enhancements.  We find that the presence of simultaneous enhancements in both the star formation and SMBH accretion rates depends on both the mass ratio of the merger and on the gas mass of the post-merger galaxy.  Despite these accretion rate enhancements, only $\sim35$\% of post-mergers experience a luminous AGN ($L_{bol}>10^{44} \mathrm{erg/s}$) within 500 Myrs after coalescence, and fewer than 10\% achieve a luminosity in excess of $L_{bol}>10^{45} \mathrm{erg/s}$. Moreover, only $\sim$10\% of the highest luminosity ($L_{bol}>10^{45} \mathrm{erg/s}$) AGN in the IllustrisTNG galaxy sample are recent mergers. Our results are therefore consistent with a picture in which mergers \textit{can} (but don't always) trigger AGN activity, but where the majority of galaxies hosting high luminosity AGN are not recent mergers.
\end{abstract}




\section{Introduction}
\label{sec:introduction}

It is well established that galaxy mergers play a fundamental role in the growth of galaxies in the Universe. In addition to the assembly of the main matter components, the merger process is predicted to have short-lived, but dramatic, effects on numerous galactic properties.  

The most obvious visible impact of a galaxy-galaxy interaction is the re-arrangement of stellar material, an effect which was seen in the earliest N-body simulations \citep[][]{Toomre1972,White1978,Roos1979,Villumsen1982}.  Strong tidal forces result in disturbed morphologies including shells and tidal features that are dependent on the properties of both the initial galaxy members and the orbital properties of the interaction \citep[][]{Casteels2014,Nevin2019,Blumenthal2020,Patton2016}. The advent of hydrodynamical simulations further demonstrated that the generation of internal asymmetric structures produced torques that could drain angular momentum from the gas, causing it to flow to the centre and resulting in central starbursts \citep[][]{Hernquist1989a,Barnes1991,Mihos1996,DiMatteo2007,Capelo2016,Blumenthal2018}, a feature that is widely observed in both the pre- \citep[][]{Barton2000,Woods2006,Woods2007,Ellison2008,Woods2010,Scudder2012,Patton2013,Knapen2015,Cao2016} and post- \citep[][]{Ellison2008,Ellison2013,Thorp2019,Bickley2022} merger regimes in galaxy surveys.

Merger driven gas inflows have also been predicted to potentially fuel black hole growth \citep[][]{Sanders1988,DiMatteo2005,Springel2005,Hopkins2008,Capelo2015}.  However, whilst there is broad agreement in the observational literature supporting merger-enhanced star formation, there continues to be controversy over the role of mergers in triggering active galactic nuclei (AGN).  Although numerous observational studies do not find evidence for a merger-AGN connection \citep[][]{Cisternas2011,Schawinski2011,Kocevski2012,Bohm2013,Shah2020,Lambrides2021}, many others have demonstrated a statistically significant excess of AGN in interacting pairs and post-mergers \citep[][]{Alonso2007,Ellison2011,Bessiere2012,Ellison2013,Hong2015,Kocevski2015,Hewlett2017,Marian2020,Bickley2022Submitted}.  

Recently, \cite{Ellison2019} explored the possibility that contradictory results may result from different experimental approaches; namely that some studies assess the excess of AGN in mergers, whereas others quantify the excess of mergers amongst AGN.  Using a sample of low redshift AGN and high quality imaging from the Canada France Imaging Survey, \cite{Ellison2019} demonstrated that there is \textit{both} an excess of AGN in mergers \textit{and} an excess of mergers amongst AGN. At least at low redshift (and for the optical and mid-IR selection used by \citealt{Ellison2019}), experimental approach does not seem to explain the conflicting results.

There are numerous other factors that may affect our assessment of the merger-AGN connection.  AGN are a multi-wavelength phenomenon that manifest across the electromagnetic spectrum and different selection methods identify different objects (see \citealt{HickoxAlexander2018} for a recent review).  The extent of a connection between mergers and AGN may depend on the wavelengths at which the AGN is selected.  For example, several studies have shown that the AGN excess in mergers is greater for mid-IR selected AGN, compared to those identified through their optical emission lines \citep[][]{Satyapal2014,Weston2017,Goulding2018,Ellison2019,Gao2020}, with no X-ray AGN enhancement found in low redshift post-mergers \citep[][]{Secrest2020}.  However, even for selection in a given frequency regime disagreement can exist.  For example, some studies of radio-selected AGN find no merger connection \citep[][]{Dicken2012,Ellison2015} whilst others do \citep[][]{RamosAlmeida2011,RamosAlmeida2012,Gao2020,Pierce2022}, a tension that could be linked to further distinctions within the radio population into high and low excitation sources \citep[][]{Chiaberge2015,Bernhard2022}.  Disagreement also exists as to whether the dominance of merger triggering is linked to AGN luminosity \citep[][]{Schawinski2012,Treister2012,Villforth2014,Glikman2015,Mechtley2016,Villforth2017,Marian2019,Pierce2022}.

The above summary of the observational literature demonstrates the considerable diversity amongst recent results.  However, there are a few broad statements regarding the merger-AGN connection that are relatively uncontroversial.  First, that the majority of AGN are \textit{not} merger induced, a statement that is also supported by simulations \citep[][]{Steinborn2018,McAlpine2020}. Second, that \textit{some} AGN are probably triggered by interactions, although the fraction likely depends on subtleties such as redshift \citep[e.g.][]{Rosario2015} and selection method \citep[e.g.][]{Gao2020}. Third, it is the most obscured \citep[][]{Satyapal2014,Fan2016,Weston2017,Ellison2019,Gao2020} and most luminous \citep[][]{Schawinski2012,Treister2012,Glikman2015,Donley2018,Goulding2018,UrbanoMayorgas2019,Pierce2022} AGN that seem to be most likely to be linked to mergers.

Beyond the simple understanding of which mechanisms (merging versus various secular processes) lead to black hole growth, the potential for AGN triggering during the interaction is arguably most relevant for assessing the end point of the merger sequence. Since AGN have been widely implicated as a way to quench star formation \citep[][]{DiMatteo2005,Springel2005}, if most mergers experience an accretion+feedback event during the interaction, then we might expect most post-coalescence galaxies to rapidly shut-down their star formation \citep[][]{Hopkins2008}.

An additional complication to the proposed connections between mergers, star formation and AGN triggering, followed by feedback driven quenching, is the different timescales for these processes. High resolution simulations have shown that merger induced central starbursts and high SMBH accretion rates are for the most part not temporally correlated \citep[][]{Volonteri20151}. Furthermore, most observational studies do not find a connection between central starburst and AGN activity, except in the highest luminosity systems, an effect that is largely attributed to short timescales for AGN variability \citep[][]{RowanRobinson1995,Schweitzer2006,Lutz2010,Shao2010,Santini2012,Rosario20131,Rosario20132}.

In order to assess the impact of mergers on star formation, black hole accretion and quenching in a statistical and holistic way, we have been undertaking a series of studies using a large, state-of-the-art hydrodynamical simulation suite from IllustrisTNG (hereafter TNG; \citealt[][]{Nelson2017,Naiman2018,Mariancci2018,Pillepich20171,Springel2017}).  Although large box simulations are lacking in resolution compared with idealized binary merger runs or cosmological zoom in simulations (e.g. \citealt{Hopkins2013,Moreno2015,Moreno2019}), they offer several advantages.  First, many thousands of interactions can be drawn from the large sample of galaxies in a cosmological simulation, in contrast to the suites of binary mergers which typically contain only tens of interactions. Second, no a priori decisions are required to define either galaxy properties (morphologies, gas fractions etc.) or interaction properties (e.g. mass ratios, orbits etc.).  Finally, the full cosmological setting provides a more realistic interaction with both the circumgalactic and intergalactic environments.

In the first paper in our series, \cite{Patton2020} demonstrated that enhancements in the star formation rate (SFR) of interacting galaxies in TNG are experienced in the pre-coalescence phase out to separations of a few hundred kpc. Although the exact number varies depending on the simulation used, triggered star formation out to large separations was found to be a common feature of all of simulations in the TNG suite. \cite{Hani2020} continued the assessment of SFR enhancement into the post-coalescence regime and showed that the elevated levels persist for typically 500 Myrs post-merger. \cite{Quai2021} demonstrated that although most post-mergers do not lead to rapid shut down of star formation, there is nonetheless an excess of quenched galaxies amongst simulated post-mergers compared with controls.

In the work presented here, we investigate the role of mergers in triggering AGN in TNG, specifically whether there is an enhancement of supermassive black hole accretion rates in the post-merger phase. In Section \ref{sec:methods}, we summarize the salient details of the TNG simulation, our merger sample selection, and our algorithm for matching post-merger galaxies to control comparisons. In Section \ref{sec:Results} we quantify the enhancement of black hole accretion rates in post-mergers and investigate whether these enhancements depend on galaxy and merger properties (Section \ref{subsec:Enhancements}). Furthermore, we quantify the timescale over which accretion rate enhancements persist (Section \ref{subsec:timescale}) and compare this with the triggered star formation in order to assess synchronicity between these processes (Section \ref{subsec:correlations}).  Finally, we address whether mergers are likely to produce AGN activity and quantify the contribution of recent mergers to the total AGN population, i.e. what fraction of mergers produce high accretion rates (Section \ref{subsec:fractionAGN}) and what fraction of high accretion rate systems have undergone a recent merger (Section \ref{subsec:fractionMergers}).  

\section{Methods}
\label{sec:methods}

\subsection{IllustrisTNG}
\label{subsec:Illustris}

We use the IllustrisTNG galaxy formation simulation, a state-of-the-art cosmological magneto-hydrodynamic simulation including galactic scale stellar feedback, stellar population evolution and chemical enrichment, primordial and metal-line gas cooling and heating, and multi-mode blackhole feedback \citep[][]{Nelson2017,Naiman2018,Mariancci2018,Pillepich20171,Springel2017}. In the work presented here, we use the TNG100-1 simulation run, the intermediate volume and resolution run of the fiducial TNG galaxy formation model, which has a $(110.7 \, \mathrm{Mpc})^3$ volume, a baryonic resolution of $1.4 \times 10^6 \, \mathrm{M_{\odot}}$, and a dark matter resolution of $7.5 \times 10^6 \, \mathrm{M_{\odot}}$.

We briefly describe the physical models for supermassive black hole seeding and feedback in TNG, with the full details available in \cite{Weinberger2017} and \cite{Pillepich2018}. Supermassive black holes, hereafter SMBHs, are seeded at $M_{BH} = 8 \times 10^5 \, \mathrm{h^{-1} M_{\odot}}$ at the centre of halos that meet a threshold mass of $5 \times 10^{10} \, \mathrm{h^{-1} M_{\odot}}$. The black holes can then grow by either accreting gas from the region surrounding the black hole, or by merging with other black holes. 

SMBH accretion is calculated using a Bondi-Hoyle-Lyttleton subgrid model,
\begin{equation}
    \dot M_{Bondi} = \frac{4 \pi G^2 M_{BH}^2 \rho}{c_s^3}
    \label{eq:Bondi}
\end{equation}
where $G$ is the gravitational constant, $M_{BH}$ is the black hole mass, and $\rho$ and $c_s$ are the density and sound speed sampled in a kernel-weighted sphere centred on the SMBH, labelled the accretion region. The SMBH accretion rate is capped by the Eddington rate, 
\begin{equation}
    \dot M_{Edd} = \frac{4 \pi G M_{BH} m_p}{\epsilon_r \sigma_T c}
    \label{eq:Edd}
\end{equation}
where $m_p$ is the proton mass, $\epsilon_r$ is the radiative accretion efficiency, $\sigma_T$ is the Thompson cross-section, and $c$ is the vacuum speed of light. 

SMBHs are merged when the black holes are within the accretion regions of one-another. Furthermore, in order to prevent the wandering of SMBHs away from the halo centre, SMBHs are fixed to the local gravitational potential minima, which has the added effect of promptly merging SMBHs when their host subhalos coalesce.

TNG uses a dual feedback mode model, where SMBHs use a different feedback prescription depending on the SMBH accretion rate and SMBH mass. At high accretion rates, feedback is implemented according to a radiative mode model, and at low accretion rates the feedback is implemented according to a kinetic mode model. The classification of high vs. low accretion is made by calculating the ratio of the Bondi accretion rate to the Eddington accretion rate, where a high accretion rate is defined as $\dot M_{Bondi} \geq \chi \dot M_{Edd}$ and $\chi$ is defined as,
\begin{equation}
    \chi = min\bigg[\chi_0 \bigg(\frac{M_{BH}}{10^8 M_{\odot}}\bigg)^{\beta},0.1\bigg].
    \label{eq:chi}
\end{equation}
The maximum $\chi=0.1$ follows observational constraints set by X-ray binaries \cite{Dunn2010}, whereas the parameters $\chi_0$ and $\beta$ are simulation parameters tuned to 0.002 and 2 respectively. 

For SMBHs with high accretion rates, the simulation uses a radiative feedback prescription, where a fraction of the accreted mass energy is injected as thermal energy into the region surrounding the black hole. At low accretion rates, once again a fraction of the accreted mass energy is injected into the surrouding BH region, however the energy injected is in the form of kinetic energy in randomized directions away from the black hole. For our work, focused on SMBH accretion rates, we do not comment in detail on the feedback prescriptions, and refer the reader to \cite{Weinberger2017}.

\subsection{Identifying galaxies and galaxy mergers in IllustrisTNG}
\label{subsec: ID mergers}

Before we can identify the population of galaxy mergers in the TNG simulation, we first apply selection constraints to our entire galaxy sample.  

First, we exclude galaxies at redshift$>$1 from our sample because the increased frequency of mergers and interactions exacerbate issues such as numerical stripping and subhalo switching \citep[][]{Rodriguez-Gomez2015}, which makes robust merger identification more difficult. We then apply a stellar mass requirement of $10^9 M_{\odot}$, equivalent to $\sim 1000$ particles, to ensure our selected galaxies are properly resolved. We combine our minimum stellar mass requirement with our minimum merger mass ratio of 1:10, i.e. requiring post-mergers have progenitors of at least $10^9 M_{\odot}$, by requiring a minimum stellar mass of $10^{10} M_{\odot}$ in the post-merger sample.

We also apply an additional environmental constraint to remove galaxies susceptible to numerical stripping from ongoing interactions with neighbouring subhalos. Following the procedure of \cite{Patton2020}, we calculate $r_{sep}$,
\begin{equation}
    r_{sep} = \frac{r}{R^{host}_{1/2} + R^{comp}_{1/2}}
    \label{eq:rsep}
\end{equation}
where r is the 3D separation of the galaxies, and $R^{host}_{1/2}$ and $R^{comp}_{1/2}$ are the stellar half mass radii of the host and companion. \cite{Patton2020} demonstrated that TNG galaxies with $r_{sep} \lesssim 2$ had stripped stellar mass and we therefore excluded them from our galaxy sample.

Once we have identified appropriate candidate galaxies from TNG, we identify galaxy mergers using the merger-trees created by the \textsc{SUBLINK} algorithm \citep[][]{Rodriguez-Gomez2015}, which associate each galaxy in TNG with progenitor and/or and descendant galaxies. We identify galaxy mergers by selecting nodes within the merger trees. Nodes occur when particles which are assigned into distinct subhalos at a given snapshot are subsequently assigned into the same subhalo in the following snapshots. 

In the work presented here, we calculate the mass ratio of the merger using a different methodology from previous works in this series. We apply additional steps, once again, as a precaution against numerical stripping and subhalo switching. For each merger, the mass ratio of the merger is calculated using the stellar mass within twice the stellar half-mass radius, in the snapshots leading up to coalescence of the subhalos, up to a maximum of 10 snapshots. From the mass ratio estimates, the maximum and minimum are excluded, which removes abnormally large or small values that are due to subhalo switching or significant numerical stripping. Finally, we take the average and standard deviation of the remaining mass ratio measurements to calculate the merger stellar mass ratio, $\mu$, and assign a merger mass ratio error, $\sigma_{\mu}$. For the work presented here, we define mergers as coalescence of subhalos with a merger mass ratio within $0.1 < \mu \leq 1$. We do not limit the overall post-merger sample using the mass ratio error except when specified. 

\subsection{Post-merger and non-merger galaxy samples}
\label{subsec: PM and NM}

\begin{figure}
	\includegraphics[width=\columnwidth]{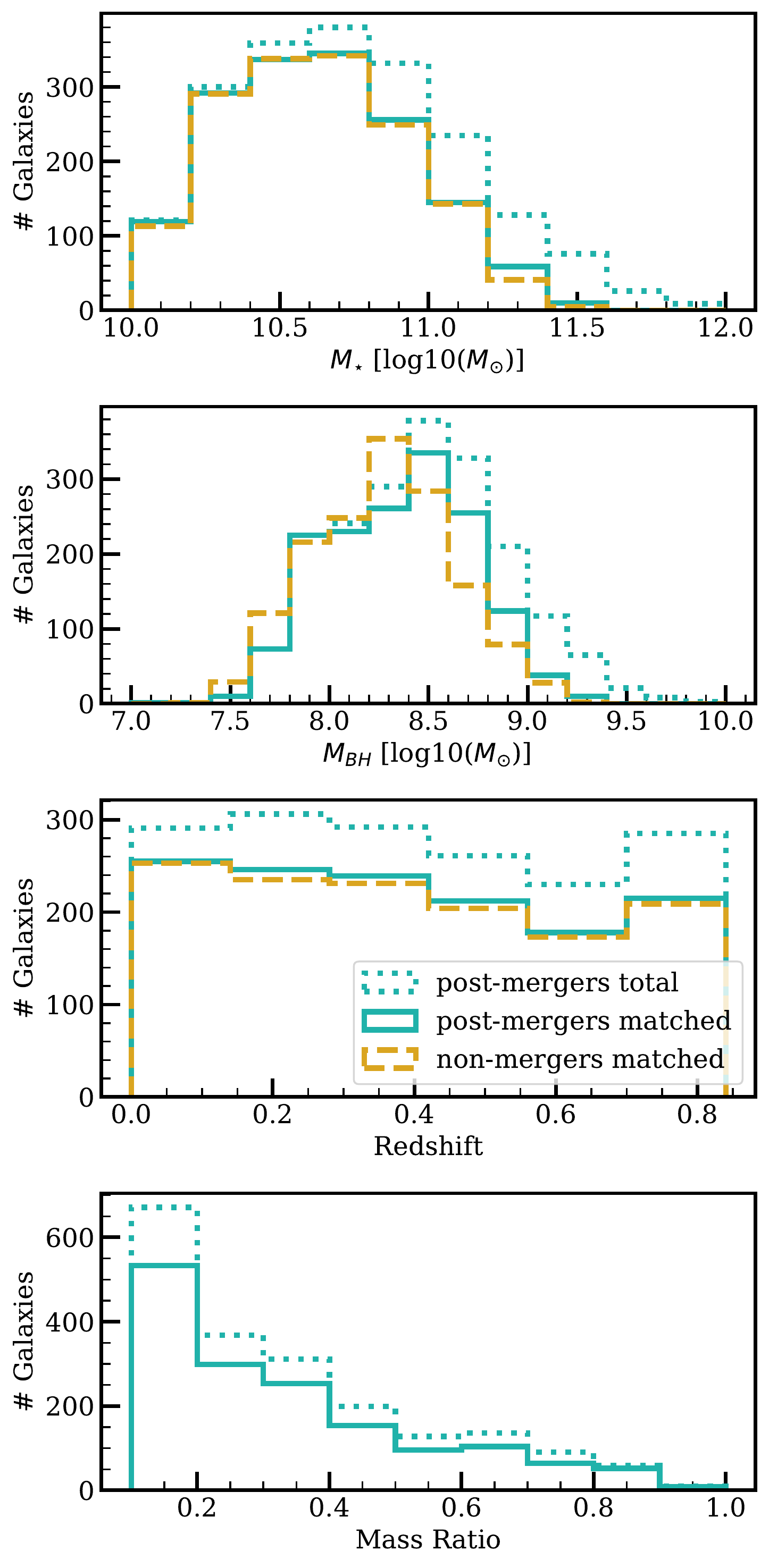}
    \caption{Distribution of post-merger and non-merger sample (from top to bottom) in stellar mass, SMBH mass, redshift, and mass ratio. The dotted teal line shows the distribution of all the post-mergers, following our criteria, in TNG100-1. The solid teal line shows the distribution of the 1563 successfully control matched post-mergers, and the dashed yellow line shows the 1522 successfully control matched non-mergers.}
    \label{fig:PMdist}
\end{figure}

Hydrodynamical simulations have suggested that SMBH accretion rates peak post-coalescence \citep[][]{Hopkins2008}, a framework that is corroborated by some observational studies \citep[][]{Ellison2013,Satyapal2014,Bickley2022Submitted}. In the work presented here we focus on post-merger galaxies, and construct our post-merger sample using only galaxy mergers that have coalesced in the time between the previous and current snapshot, or within $\sim 160$ Myrs. Note that we discuss the long term effects on the post-merger sample in Section \ref{subsec:timescale}. Selecting galaxies immediately after coalescence, and applying the galaxy selection criteria outlined above, we identify a sample of 1973 post-merger galaxies from TNG100-1. The stellar mass, SMBH mass, redshift and mass ratio distributions of this complete post-merger sample are shown in the dotted teal line in Figure \ref{fig:PMdist}.

In addition to the post-mergers, we define a sample of non-mergers whose properties are representative of the post-mergers. We construct a non-merger sample in order to generate a comparative population that has the same underlying distribution as the post-merger sample, and in order to contextualize how the underlying stochasticity of SMBH accretion rates affect our calculation of accretion rate enhancements. The representative non-merger sample is constructed as follows. For each of the 1973 post-mergers, we identify one non-merging galaxy that is selected to be the single best match in redshift, stellar mass, gas mass, environment, and feedback mode (see Section \ref{subsec:matching} for details on the necessity of feedback mode matching) that has not undergone a merger of mass ratio $\mu > 0.1$ within the last two Gyrs. The environment is quantified using two parameters taken from \cite{Patton2020}: $r_1$ and $N_2$. $r_1$ is the 3D distance to the nearest neighbour, within two Mpc, which has a mass above 10\% of the target galaxy mass. $N_2$ is the number of neighbours within two Mpc of the galaxy centre, with a minimum mass of $10^{9} M_{\odot}$. The single best match is selected as an exact match in redshift and feedback mode, and the best simultaneous match of stellar mass, gas mass, $r_{1}$ and $N_{2}$ following the statistical weighting scheme of \cite{Patton2016}. We therefore begin with a sample
of 1973 post-mergers and 1973 non-mergers. 

\subsection{Control Matching Post-merger and Non-merger Galaxies}
\label{subsec:matching}
In order to quantify the difference between a given physical property (e.g. SMBH accretion rate) of an individual (post-merger or non-merger) galaxy and the expected value of a `normal' galaxy, we use a control matching algorithm to identify a sample of comparative controls for each individual galaxy in the post-merger and non-merger sample. The pool of possible control galaxies consists of all galaxies, meeting the criteria outlined in Section \ref{subsec: ID mergers}, that have not undergone a merger of mass ratio $\mu > 0.1$ within the last two Gyrs. We note that when control matching the non-mergers, we exclude galaxies that have already been assigned into the non-merger sample from the control pool, such that no galaxy can be matched to itself. 

\begin{figure}
	\includegraphics[width=\columnwidth]{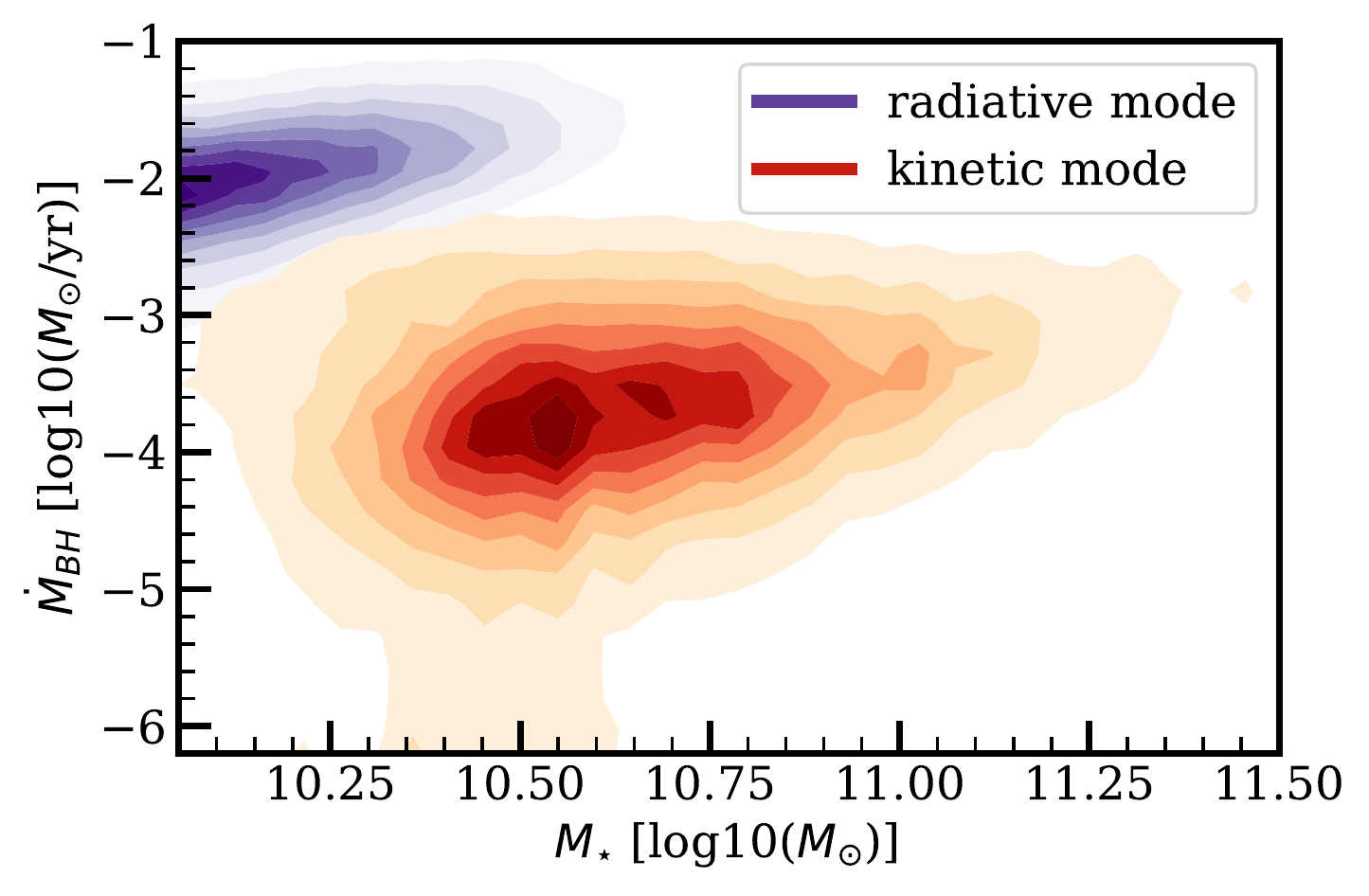}
    \caption{SMBH accretion rates as a function of stellar mass for all the TNG100-1 galaxies meeting the selection criteria of Section \ref{subsec: ID mergers}. The dark blue contours represent the sample in radiative mode feedback and the orange contours represent the sample in kinetic mode feedback.}
    \label{fig:Mdot_vs_Mstar}
\end{figure}

We begin by down-selecting the pool of potential control galaxies by imposing two cuts (which are effectively broad matching constraints). First, we require that the controls for a given galaxy are drawn from the same simulation snapshot, which corresponds to matching in redshift. Second, we require that the controls for a given galaxy are in the same instantaneous feedback mode as the post-merger (or non-merger), using the feedback mode classification outlined in Section \ref{subsec:Illustris}.  The requirement of feedback mode matching diverges from the methods applied by \cite{Patton2020}, \cite{Hani2020}, and \cite{Quai2021}, but is necessary for the work presented here to avoid spurious features in the results. The issue is illustrated in Figure \ref{fig:Mdot_vs_Mstar} which shows the distribution of accretion rates as a function of stellar mass for all galaxies in TNG100-1, colour-coded to distinguish those in radiative feedback mode (dark blue) from those in kinetic feedback mode (red). Figure \ref{fig:Mdot_vs_Mstar} demonstrates that (at a fixed stellar mass) the accretion rate distributions are bimodal. Without feedback mode matching, post-merger galaxies may (artificially) exhibit extremely enhanced or suppressed accretion rates if they are matched to controls in the other feedback mode.

In addition to snapshot (redshift) and feedback mode, the control galaxies for each post-merger and non-merger are matched on several further properties. First, the control sample is limited to galaxies whose stellar mass is matched to within $\pm 0.05$ dex and whose gas mass is matched to within $\pm 0.1$ dex. Next, the matched control sample is limited to galaxies within $\pm 10\%$ of $r_1$ and $N_2$. We allow all of the error tolerances to grow, 0.01 dex in stellar mass, 0.1 dex in gas mass, and $10\%$ in $r_1$ and $N_2$, up to four times until at least five matched control galaxies are found. If fewer than five control galaxies are found, then the post-merger is excluded from the sample. On average, galaxies have to grow their error tolerances twice to meet the required number of controls, and 410 post-merger galaxies (and 451 non-merger galaxies) did not meet the control requirements within permitted error tolerances after the maximum number of grows and were hence rejected from our sample. Therefore, our final matched sample consists of 1563 matched post-merger galaxies and 1522 non-merger galaxies, with an average of 10 matched controls each. The properties of the final post-merger and non-merger samples are shown in the solid teal and dashed yellow lines in Figure \ref{fig:PMdist}. We note that the top panel of Figure \ref{fig:PMdist} reveals that the control matching algorithm preferentially fails to match higher stellar mass post-mergers. It was determined that the strict error tolerance on stellar mass matching combined with gas mass matching resulted in higher mass post-mergers being less likely to find a sufficient number of controls. However, we note that both matching criteria are necessary in order to select an unbiased set of controls. Without the strict error tolerance, high mass post-mergers are preferentially matched to low mass controls. We therefore prioritize the selection of unbiased controls over the completeness of the matched post-merger sample.

\begin{figure}
	\includegraphics[width=\columnwidth]{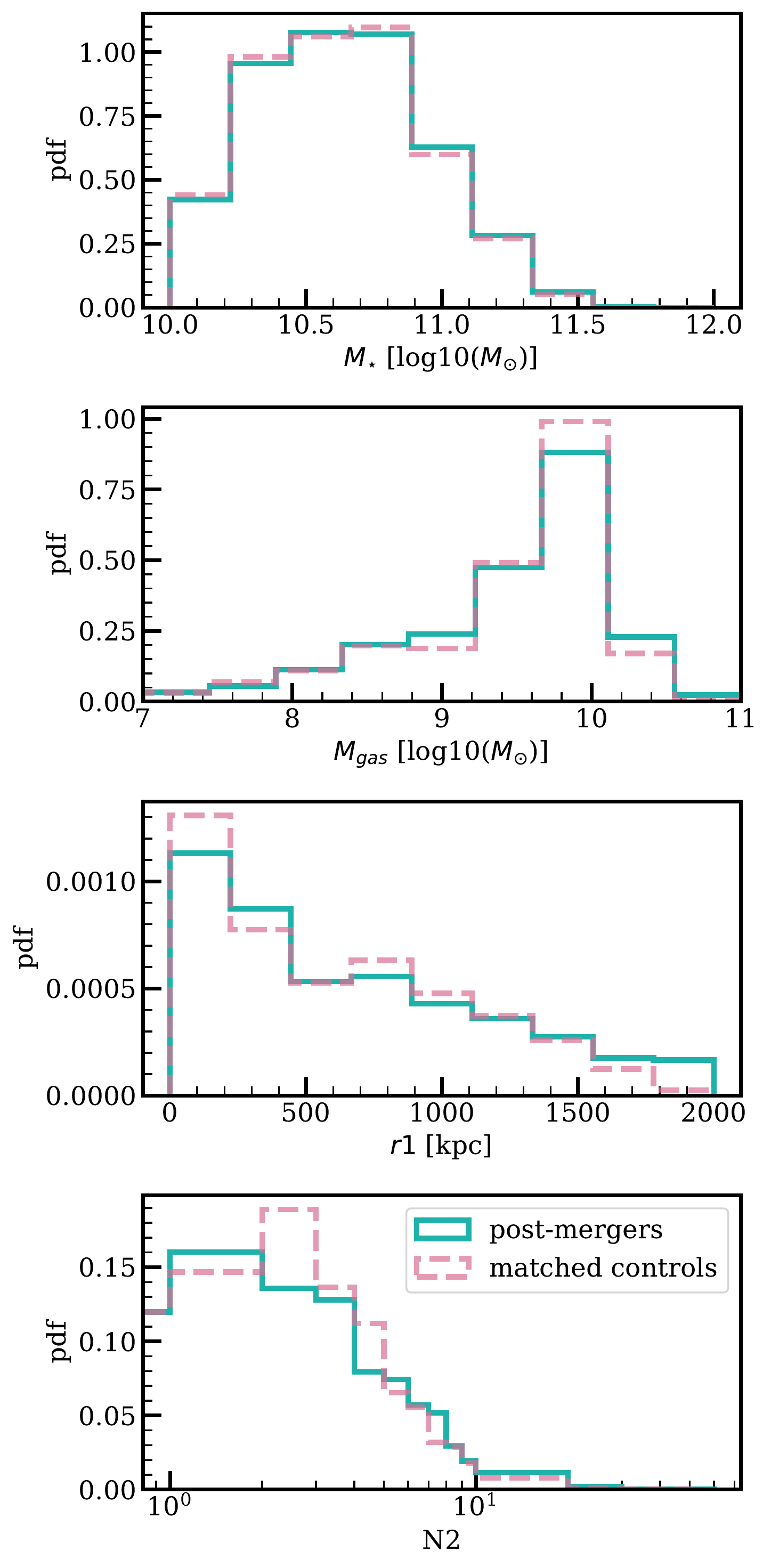}
    \caption{Distribution of parameters matched in control matching process: (from top to bottom) stellar mass and gas mass measured within twice the stellar half mass radius, distance to nearest neighbour within two Mpc, and number of neighbours within two Mpc. Post-mergers are represented with the teal line, and the controls are represented with the dashed pink line.}
    \label{fig:ControlsMatch}
\end{figure}

Figure \ref{fig:ControlsMatch} shows the distribution of stellar mass, gas mass, $r_1$, and $N_2$ for the post-merger galaxies, shown in the teal line, and the matched control galaxies, shown in dashed pink line. In the figure, the distributions trace one another closely, demonstrating the success of the control matching methodology in identifying galaxies with similar properties to the post-merger sample.  Although not shown in Figure \ref{fig:ControlsMatch}, we note that the non-merger sample and its controls exhibit similarly closely matched properties.

Once the control matching procedure is complete, we then calculate a relative SMBH accretion rate enhancement, $\Delta \dot M_{BH}$, for each post-merger and non-merger galaxy as

\begin{equation}
    \Delta \dot M_{BH} = log_{10}(\dot M_{BH}) - median[log_{10}(\dot M_{BH Controls})].
    \label{eq:enhancement}
\end{equation}

\section{Results}
\label{sec:Results}

\subsection{Enhancement of accretion rates in post-merger galaxies}
\label{subsec:Enhancements}

\begin{figure}
	\includegraphics[width=\columnwidth]{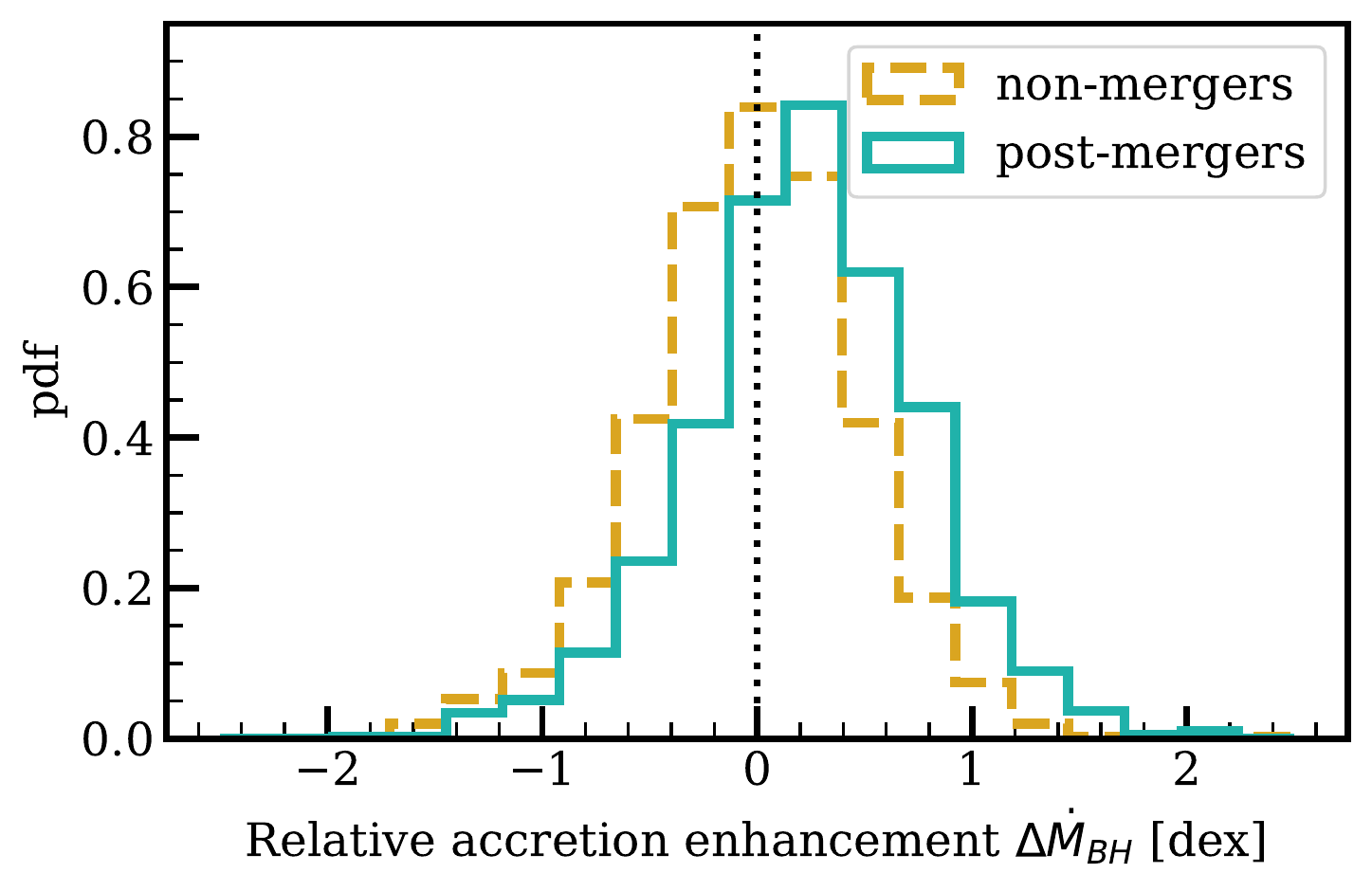}
    \caption{Histogram of accretion rate enhancements for the post-merger population and non-merger population. Post-mergers are represented with the teal line, and non-mergers are represented with the dashed yellow line. The median enhancement of the post-merger sample is 0.23 dex.}
    \label{fig:DMdot_hist_all}
\end{figure}

Figure \ref{fig:DMdot_hist_all} shows the distribution of $\Delta \dot M_{BH}$ for the post-merger sample, shown in the teal line, and non-merger sample, shown in the yellow dashed line. The non-merger population peaks at 0 dex, consistent with no enhancement in the SMBH accretion rate of non-mergers relative to their controls. A lack of statistical offset may be expected for the non-merger sample, as it is a subset of the control population. However, since the non-merger sample is not randomly selected from the control pool, and is selected to match the population characteristics of the post-merger sample, we have therefore demonstrated that the population characteristics of the non-merger sample do not give rise to enhanced SMBH accretion rates relative to their controls. Nonetheless, the non-merger distribution provides a useful reference. For example, both populations show a similar distribution of $\Delta \dot M_{BH}$, ranging from 100 times enhanced to 100 times suppressed accretion rates relative to controls, with 30 non-mergers (2\%) achieving accretion rates at least 10 times higher than controls. The comparable spread demonstrates the large variability of accretion rate in both samples. In contrast to the non-mergers, the post-merger sample peaks at a positive accretion rate enhancement, with a median enhancement of 0.23 dex, corresponding to an enhancement of $\sim 70 \%$. In the post-merger sample, 102 post-mergers (6.5\%) have accretion rates at least 10 times higher than controls. Figure \ref{fig:DMdot_hist_all} therefore demonstrates our first main result, that mergers in TNG, on average, have an elevated accretion rate in the post-merger phase.

\begin{figure}
	\includegraphics[width=\columnwidth]{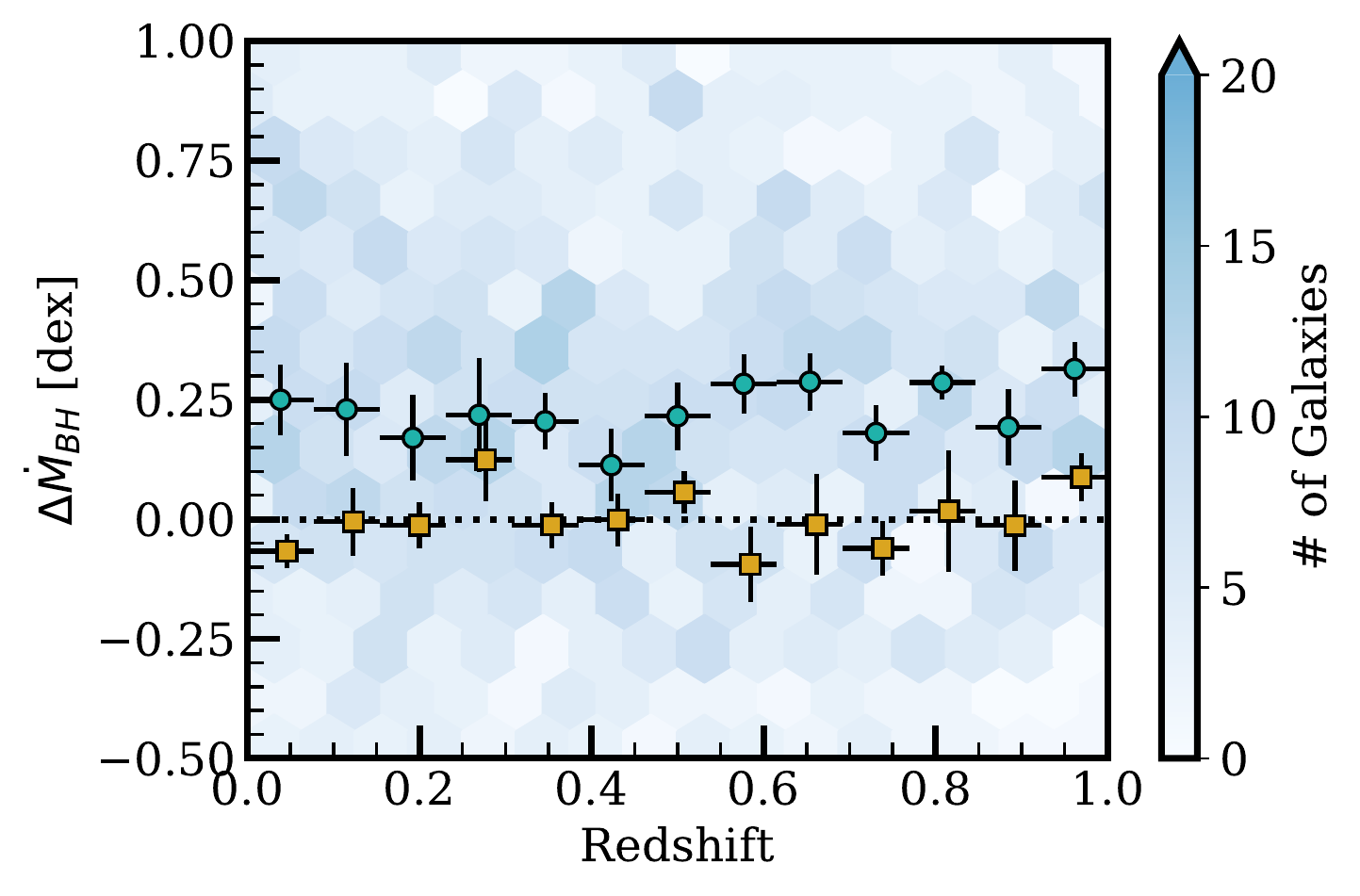}
    \caption{Accretion rate enhancements of post-merger galaxies for different redshifts. The background density plot shows the distribution of $\Delta \dot M_{BH}$ for the post-merger sample, and the foreground points represent the median value at that redshift. The teal circles are the median post-merger $\Delta \dot M_{BH}$ and the yellow squares are the median non-merger $\Delta \dot M_{BH}$. The error on the x-axis represents the bin width for each data point, and the error on the y-axis is the standard error on the median for that redshift bin.}
    \label{fig:DMdot_vs_z}
\end{figure}

Figure \ref{fig:DMdot_vs_z} shows the SMBH accretion rate enhancement of the post-merger and non-merger samples as a function of redshift. The full distribution of the post-merger sample is shown in the background density plot, and the foreground points are the median accretion enhancement for the post-merger sample (teal circles) and non-merger sample (yellow squares) within equally spaced bins of redshift. The errorbars represent the standard error on the median within each bin. The median accretion rate enhancement of the post-merger sample is consistently above 0 across the redshift range. We present both the median enhancement points and the background distribution to emphasize that while there is a positive median $\Delta \dot M_{BH}$ in post-mergers, not all post-mergers show accretion rate enhancements (as expected from Figure \ref{fig:DMdot_hist_all}). Overall, neither sample (post-mergers nor non-mergers) show any significant dependence on redshift, indicating that the merger process elevates accretion rates out to at least $z=1$ (the limit of our sample selection).

\begin{figure}
	\includegraphics[width=\columnwidth]{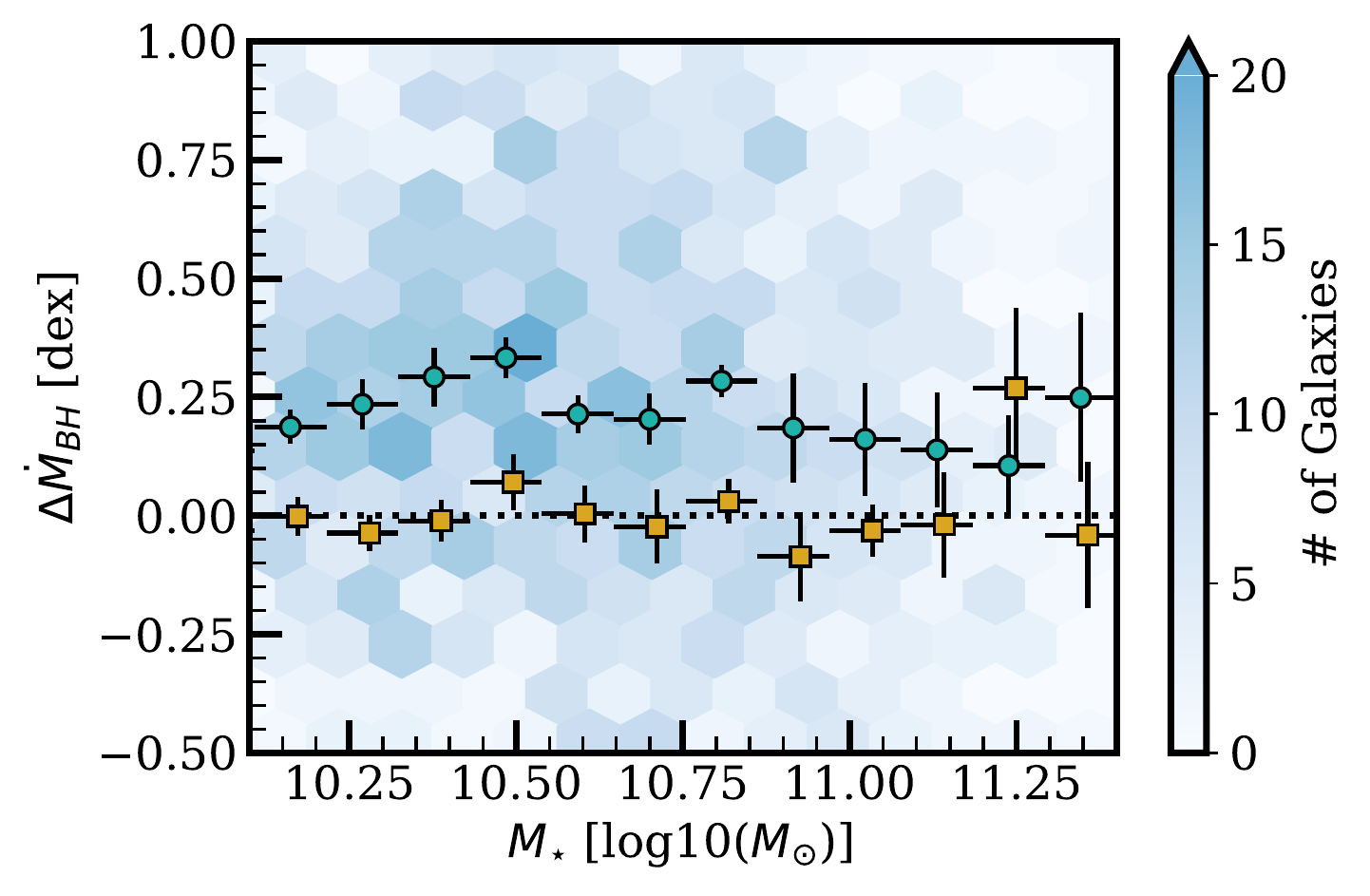}
    \caption{Accretion rate enhancements of post-merger galaxies for different stellar masses. The background density plot shows the distribution of $\Delta \dot M_{BH}$ for the post-merger sample, and the foreground points represent the median value at that stellar mass. The teal circles are the median post-merger $\Delta \dot M_{BH}$ and the yellow squares are the median non-merger $\Delta \dot M_{BH}$. The error on the x-axis represents the bin width for each data point, and the error on the y-axis is the standard error on the median for that stellar mass bin.}
    \label{fig:DMdot_vs_Mstar}
\end{figure}

We find a similarly consistent result when investigating the dependence of SMBH accretion rate enhancements on stellar mass, shown in Figure \ref{fig:DMdot_vs_Mstar}. As expected from its construction, we find that the non-merger sample has accretion rate enhancements consistent with zero at all stellar masses. In contrast the teal points demonstrate that the SMBH accretion rates in post-mergers are consistently enhanced, on average, at all stellar masses within our sample. 

\begin{figure}
	\includegraphics[width=\columnwidth]{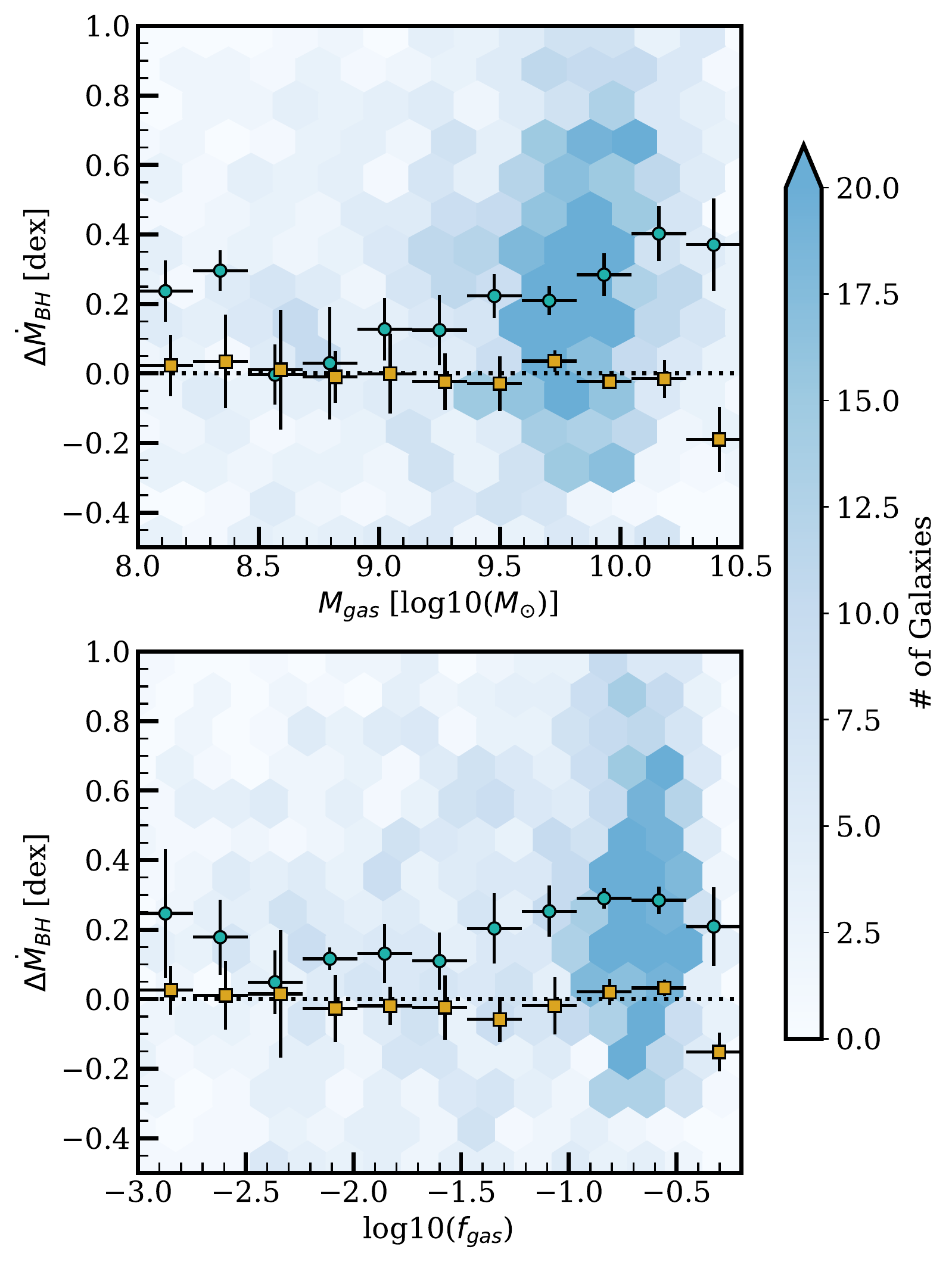}
    \caption{Accretion rate enhancements of post-merger galaxies for different $M_{gas }$ (shown in top panel) and gas fractions (shown in bottom panel). The gas mass is the sum of all gas particles within twice the stellar half-mass radius of the galaxy. The gas fraction is the ratio of the gas mass to the baryon mass (gas mass + stellar mass). The background density plot shows the distribution of $\Delta \dot M_{BH}$ for the post-merger sample, and the foreground points represent the median value at that $M_{gas}$ or gas fraction. The teal circles are the median post-merger $\Delta \dot M_{BH}$ and the yellow squares are the median non-merger $\Delta \dot M_{BH}$. The error on the x-axis represents the bin width for each data point, and the error on the y-axis is the standard error on the median for that $M_{gas}$ or gas fraction bin.}
    \label{fig:DMdot_vs_Mgas}
\end{figure}

Figure \ref{fig:DMdot_vs_Mgas} shows the accretion rate enhancements $\Delta \dot M_{BH}$ as a function of the gas mass (top panel) and gas fraction (bottom panel) of the post-merger or non-merger galaxy, where the gas mass is the sum of all gas particles within twice the stellar half-mass radius and the gas fraction is the ratio of the gas mass to the baryon mass (gas + stars). In the top panel, we see that for a gas mass less than $\sim 10^9 \mathrm{M_{\odot }}$, the median enhancements of the post-merger sample are not significantly distinct from the non-merger sample, with the exception of one point at $\sim 10^{8.3-8.4} \mathrm{M_{\odot}}$. However in the regime of $> 10^{9} \mathrm{M_{\odot}}$, post-mergers consistently have on average enhanced accretion rates. Therefore, Figure \ref{fig:DMdot_vs_Mgas} suggests that post-mergers of a lower gas mass are less likely to have enhanced accretion rates. In addition, at gas masses above $10^9 \mathrm{M_{\odot}}$, post-merger galaxies consistently show accretion rate enhancements, where non-merger galaxies of this mass range do not, suggesting that the presence of a large amount of gas does not guarantee higher than average accretion rates.

We find a qualitatively consistent result looking at the gas fraction (bottom panel of Figure \ref{fig:DMdot_vs_Mgas}). We see that for post-mergers with a lower gas fraction, there is a trend towards low to no relative accretion enhancement. Once again, there is some evidence for an exception to this trend at the lowest gas fraction, however we caution that the poor statistics at lower gas mass and gas fraction make the median data point more susceptible to the variability on a galaxy by galaxy basis. Therefore, we find that the presence of a significant amount of gas is an essential, but insufficient criterion to produce accretion rate enhancements, and that the merger itself is important for driving gas into the region surrounding the SMBH whence it can accrete. 

\begin{figure}
	\includegraphics[width=\columnwidth]{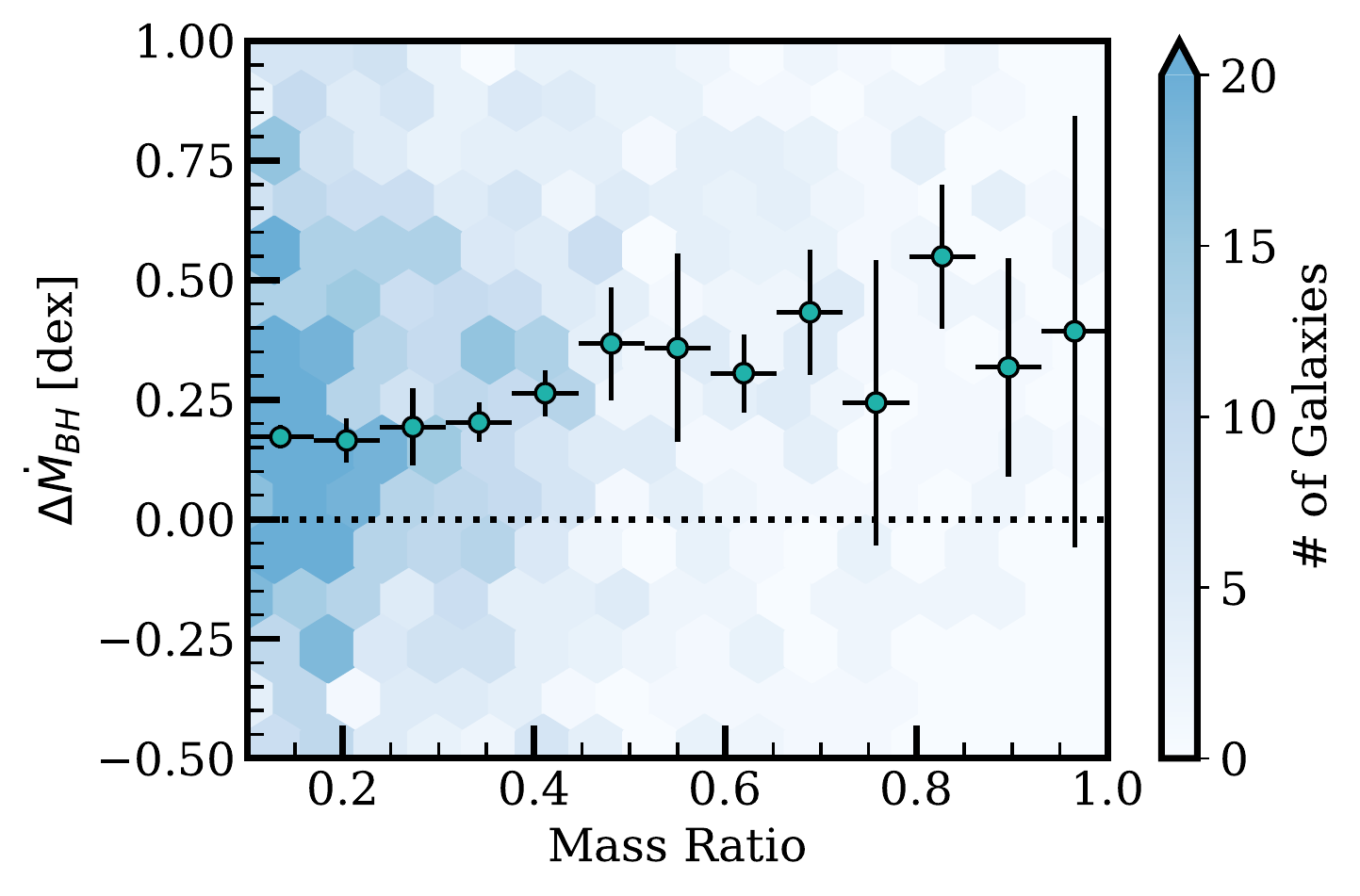}
    \caption{Accretion rate enhancements of post-merger galaxies for different merger mass ratio. The background density plot shows the distribution of $\Delta \dot M_{BH}$ for the post-merger sample, and the foreground points represent the median value at that mass ratio. The error on the x-axis represents the bin width for each data point, and the error on the y-axis is the standard error on the median for that mass ratio bin.}
    \label{fig:DMdot_vs_MassRatio}
\end{figure}

Finally, Figure \ref{fig:DMdot_vs_MassRatio} shows $\Delta \dot M_{BH}$ for post-merger galaxies of different merger mass ratios. We note that in this plot, we only include galaxies with a merger mass ratio error of $\sigma_{\mu} \leq 0.1$, which excludes 99 galaxies from the sample of 1563 post-mergers. By visual inspection, we find no significant relationship between the strength of the accretion rate enhancement and the merger mass ratio. In fact, we see that the majority of enhancements are consistent, within error, with the overall sample accretion rate enhancement of 0.23 dex. We further confirm the lack of correlation with a statistical Pearson correlation test, which yields a correlation coefficient of $\sim$0.1. 

The lack of correlation between $\Delta \dot M_{BH}$ and mass ratio is somewhat contradictory to previous simulation results. For example, \cite{Capelo2015} find mass ratio to be the most important factor influencing SMBH accretion rate in a suite of high resolution binary merger simulations. However, there are notable differences between the experiments of \cite{Capelo2015} and the work presented here. Specifically, \cite{Capelo2015} look at the effect of varying the merger mass ratio while keeping the orbital geometry and gas fraction constant, whereas our result looks at a population averaged enhancement and is therefore subject to the variable conditions of every merger. In addition, \cite{Capelo2015} comment on the role of resolution in their result, and that some of the torques generated in the galaxy interaction require high resolution simulations in order to be resolved. Finally, our result should not be interpreted as mass ratio having zero role in regulating gas inflows. Instead, Figure \ref{fig:DMdot_vs_MassRatio} demonstrates that, once the full demographic of merger properties is sampled, mass ratio is not a dominant factor in predicting the strength of a SMBH accretion rate enhancement. 

\subsection{Timescale of accretion rate enhancements}
\label{subsec:timescale}

We have thus far demonstrated that post-merger galaxies show, on average, enhanced accretion rates when compared to matched controls. This result is in agreement with previous theoretical studies that demonstrate merger driven gas inflows \citep[][]{DiMatteo2005,Springel2005,Hopkins2008}, which would increase the local gas density surrounding the black hole. However, the timescales of enhanced SMBH accretion are crucial to understanding the influence of the merger beyond the immediate post-merger phase, and the role of merger induced accretion rate enhancements on a galaxy's evolution. In addition, studying the average temporal behaviour of SMBH accretion rate enhancements can help alleviate the challenges associated with studying the highly stochastic individual SMBH accretion rates.

In the following section, we re-perform an experiment from a previous paper in this series. Following the procedure of \cite{Hani2020} (who investigated enhancements in SFR), we calculate a $\Delta \dot M_{BH}$ for post-merger galaxies in the snapshots following coalescence. Specifically, for each post-merger in our sample, we follow the descendants for as long as the merger tree allows or until the next merger event (with a mass ratio greater than 1:10). We control match the post-merger descendants to identify the enhancement to SMBH accretion rate as a function of the time since the most recent merger. We re-select the control galaxies at each subsequent snapshot following the same matching criteria as Section \ref{subsec:matching}. Therefore, a galaxy at snapshot N will not be matched to the same controls in the subsequent snapshot N+1 if the controls fall out of the acceptable matching criteria and tolerances specified in Section \ref{subsec:matching}.

\begin{figure}
	\includegraphics[width=\columnwidth]{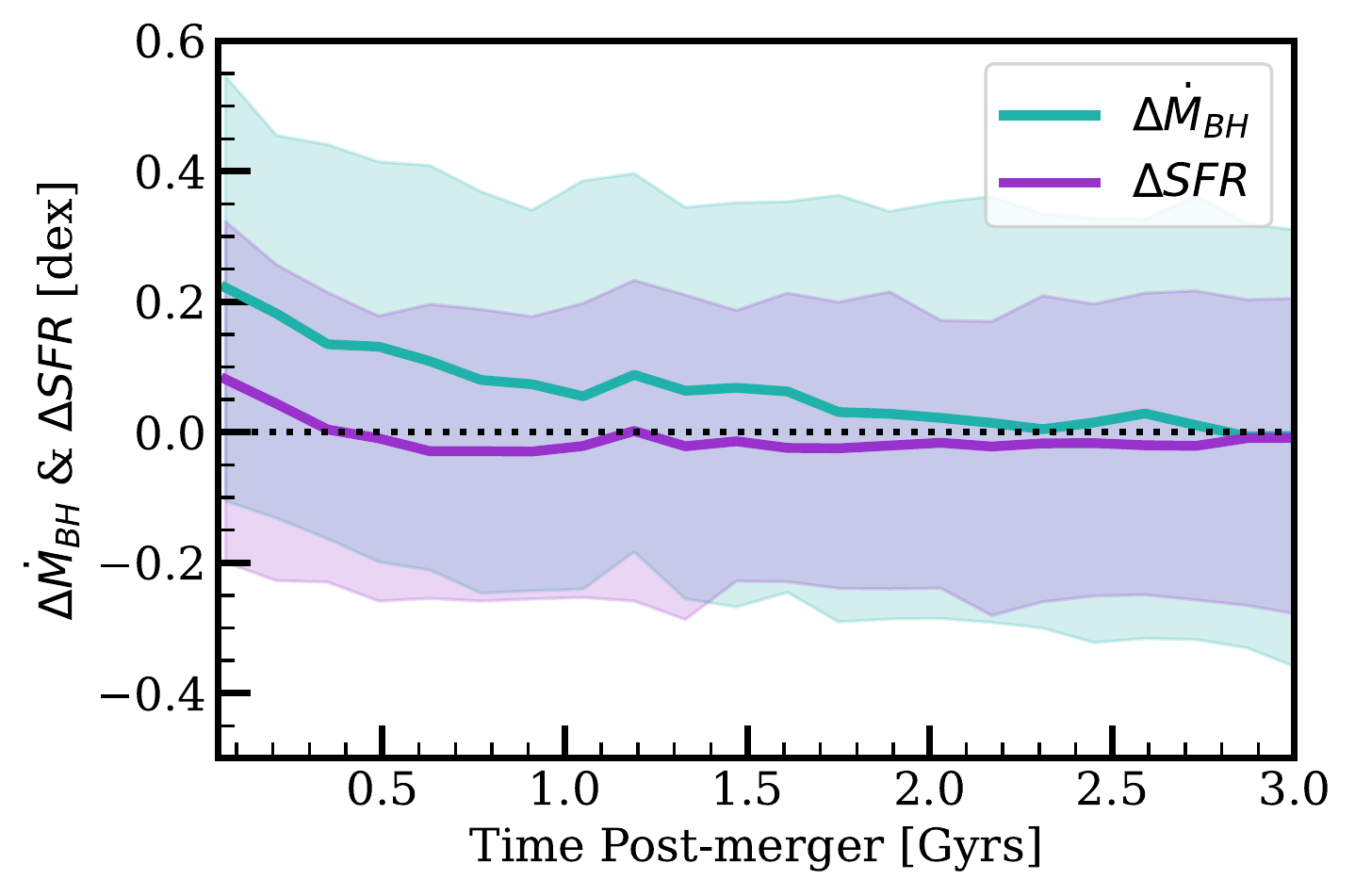}
    \caption{Running median of the accretion rate enhancement (teal) and star formation rate enhancement (purple) over the time post-merger. The error of the shaded region represents the 25th and 75th percentiles associated with the median at that time post-merger.}
    \label{fig:DMdot_DSFR_TPM}
\end{figure}

Figure \ref{fig:DMdot_DSFR_TPM} shows the SMBH accretion rate enhancement as a function of time post-merger. The median SMBH accretion rate enhancement is shown in the teal line, where the shaded region represents the 25th and 75th percentiles. We find that, on average, SMBH accretion rate enhancements persist for up to two Gyr post-merger. A long-lived accretion rate enhancement may seem to be in contradiction with the short lived (tens-hundreds of Myrs) high accretion rate events observed in previous simulation studies \citep[][]{DiMatteo2005,Springel2005,Hopkins2008}. However we emphasise our relative enhancement variable $\Delta \dot M_{BH}$ is distinct from an enhancement in absolute accretion rate in a given galaxy. Our result demonstrates that post-mergers sustain a population-averaged enhanced accretion rate relative to matched controls, which does not necessarily correspond to a long lived high accretion rate event in an individual galaxy. A long-lived population average suggests that the dynamical disturbance that leads to higher than normal feeding of the SMBH persists for up to $\sim 2$ Gyrs in TNG100-1.

We are also interested in investigating the temporal correlation, or lack there-of, between star formation rate enhancements and SMBH accretion rate enhancements. Studying the population averaged enhancements can smooth out the large temporal variability of SMBH accretion rates, and allow us to look at the timescales over which connections between star formation and SMBH accretion rate are present in the overall post-merger population. Therefore, we also compute the star formation rate enhancements of our post-merger sample,

\begin{equation}
    \Delta SFR = log_{10}(SFR) - median[log_{10}(SFR_{Controls})],
    \label{eq:enhancement2}
\end{equation}

where the star formation rate is the sum of star formation rates for all cells within twice the stellar half-mass radius. When calculating $\Delta SFR$, we introduce an additional matching criterion from \cite{Hani2020}. We require that galaxies be matched within a classification of star forming or passive in order to avoid (spuriously) large/small values of $\Delta SFR$. Passive galaxies are defined as galaxies which lie more than 2$\sigma$ below the star forming main sequence, where we calculate the star forming main sequence by applying a linear fit to TNG100-1 galaxies with stellar mass between $10^{9-10.2}M_{\odot}$ and between redshift 0 to 1, and extrapolate the linear fit to higher stellar masses following the procedure of \citep[][]{Donnari2019}. In this way, post-merger star-forming galaxies are matched to star-forming controls, in addition to the fiducial matching criteria outlined in Section \ref{subsec:matching}.  Likewise, passive post-mergers are matched to passive controls. We therefore identify a different set of controls when calculating $\Delta \dot M_{BH}$ and $\Delta SFR$ for the experiments presented in Section \ref{subsec:timescale}. 

Figure \ref{fig:DMdot_DSFR_TPM} shows the star formation rate enhancement in the purple line, once again the shaded region represents the 25th and 75th percentiles. We qualitatively recover the result of \cite{Hani2020}, who find that, on average, post-merger galaxies demonstrated star formation rate enhancements\footnote{We note that the magnitude of the star formation rate enhancement in post-merger galaxies is lower in the work presented compared with the main result of \cite{Hani2020}. There exist a number of subtle matching scheme differences, specifically the inclusion of gas mass matching (which is demonstrated in \cite{Hani2020} to reduce the magnitude of the SFR enhancement).}, and that the enhancements persisted for up to $\sim 500$ Myrs after the merger, consistent with previous simulation studies \citep[][]{DiMatteo2008} and observational estimates \citep[][]{Wild2010}. We therefore demonstrate that, on average, SMBH accretion rate enhancements are significantly longer lived than star formation rate enhancements within our post-merger sample. We compare the result of Figure \ref{fig:DMdot_DSFR_TPM} with \cite{Volonteri20151}, who investigate the temporal correlation between SFR and SMBH accretion rate in 10 high resolution binary merger simulations. \cite{Volonteri20151} find that, on an individual merger basis, properties are only temporally correlated within the 200-300 Myrs of the final coalesence event of the merger, a timescale that is consistent with our population averaged result. Our result that enhancements of SMBH accretion rates can be sustained up to 2 Gyrs is also consistent with the result of \cite{Volonteri20152}, who find that the SMBH accretion rates of merger remnants can remain sufficiently high such that the luminosity of the AGN is dominant over the stellar luminosity up to 1.5 Gyrs after the merger.

As a possible explanation for the difference in the timescale of SFR and accretion rate enhancements, we comment that in TNG, both the star formation and SMBH accretion depend on local gas density, where stars form in gas following the empirical Kennicutt-Schmidt relation \citep[][]{Schmidt1959,Kennicutt1998} and the gas dependence of SMBH accretion is shown in Eq. \ref{eq:Bondi}. However, star formation in TNG additionally requires a threshold density of $n_{H} \gtrsim 0.1 cm^{-3}$ \citep[][]{Pillepich2018}. We suggest a possible scenario where a merger event increases the central gas density significantly within $\sim 500$ Myrs of coalesence, resulting in enhanced star formation and SMBH accretion rates. However, the increase of material to the central gas reservoir is insufficient to sustain star formation past 500 Myrs yet sufficient to sustain a long lived low accretion rate enhancements.

We considered whether the above average accretion rates in the post-merger sample, which persist $\sim$ two Gyrs after the merger event, may be a feature of the matching methodology, as we define controls as galaxies that have not undergone a merger of mass ratio $>1:10$ within the last two Gyrs. We test whether the timescale of averaged accretion rate enhancements is sensitive to the minimum elapsed time post-merger that we allow for controls. We regenerate our control sample, now requiring that control galaxies to have had at least three Gyrs since their most recent 1:10 merger, and find that the timescale of two Gyrs is robust.

Figure \ref{fig:DMdot_DSFR_TPM} demonstrates the averaged behaviour of $\Delta \dot M_{BH}$ and $\Delta SFR$ in the total post-merger sample. However, not all galaxies show positive SMBH accretion rate or SFR enhancements (e.g. Figure \ref{fig:DMdot_hist_all}). In particular, the results of Section \ref{subsec:Enhancements} suggest that gas poorer galaxies do not exhibit, on average, accretion rate enhancements. We therefore separate the post-merger population based on the instantaneous feedback mode, which separates the post-mergers with high absolute accretion rates (radiative mode) from low absolute accretion rates (kinetic mode). The distinction in feedback mode also broadly separates lower mass, gas rich galaxies (radiative mode) from higher mass, gas poor galaxies (kinetic mode). 

\begin{figure}
	\includegraphics[width=\columnwidth]{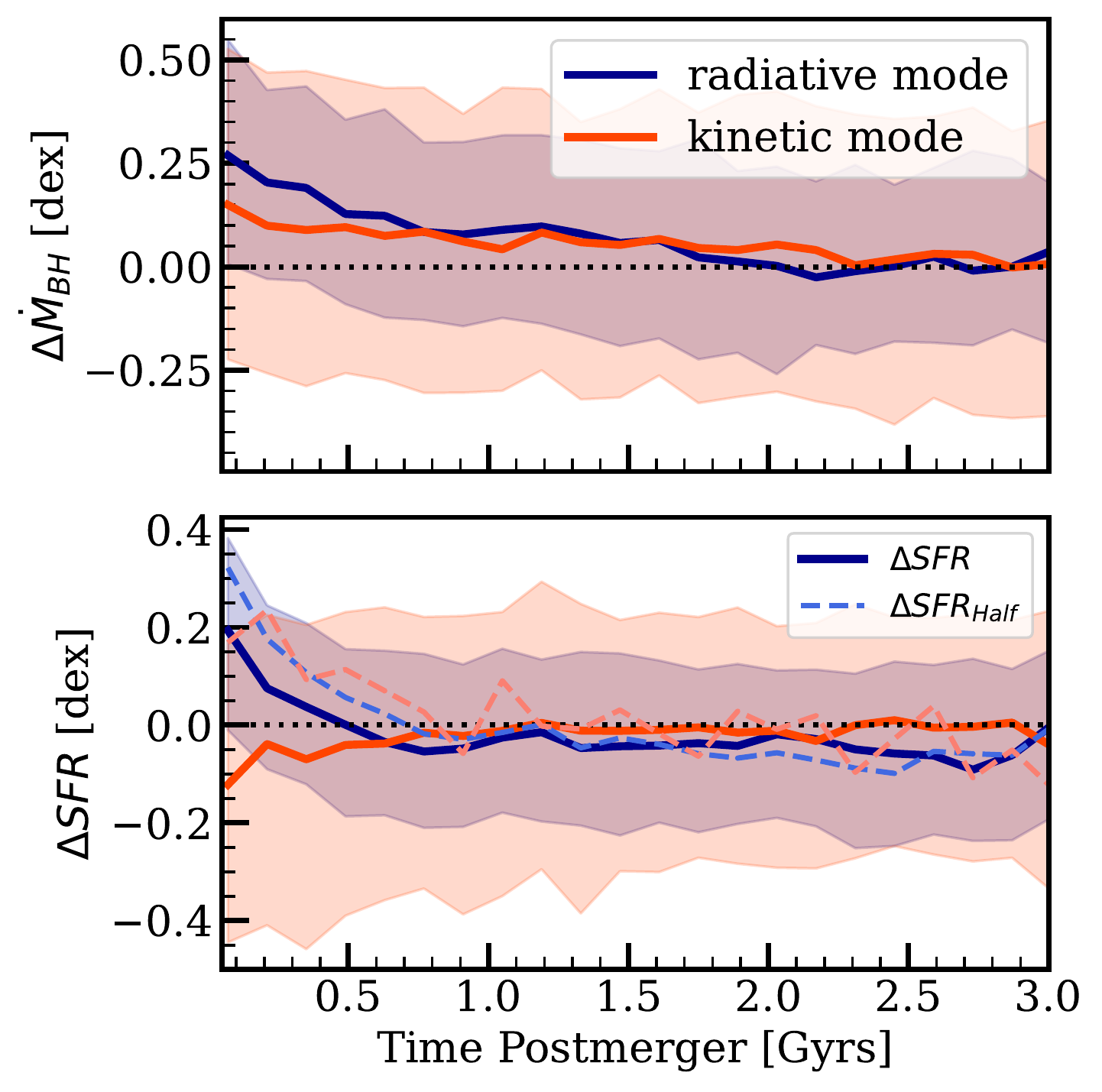}
    \caption{Running median of the accretion rate enhancement (top panel) and the star formation rate enhancement (bottom panel) over the time post-merger. The dark blue line represents the population in radiative mode feedback, and the orange line represents the population in kinetic mode feedback. The solid lines show the star formation rate measured within twice the stellar half-mass radius. The dashed lines represent the star formation rate enhancement calculated in a smaller aperture of one stellar half-mass radius. The error on the y-axis represents the 25th and 75th percentiles associated with the median $\Delta \dot M_{BH}$ and $\Delta SFR$ at that time post-merger.}
    \label{fig:DMdot_DSFR_TPM_split}
\end{figure}

Figure \ref{fig:DMdot_DSFR_TPM_split} shows SMBH accretion rate enhancements and SFR enhancements, separated into the radiative mode, shown in blue, and kinetic mode, shown in red. In the top panel, we see that both populations show, on average, positive accretion rate enhancements within two Gyrs of coalescence. The radiative mode feedback population has a higher peak accretion rate enhancement of $\sim$ 0.3 dex compared with a relative enhancement of $\sim$0.2 dex in the kinetic mode feedback population. The lower median accretion rate enhancements in kinetic mode feedback galaxies is consistent with Section \ref{subsec:Enhancements}, demonstrating that accretion rate enhancements are lower in galaxies with lower gas mass or gas fraction. 

The bottom panel of Figure \ref{fig:DMdot_DSFR_TPM_split} shows the star formation rate enhancement separated by feedback mode, where the solid lines show the star formation rate enhancement calculated within twice the stellar half-mass radius and dashed lines within one stellar half-mass radius. We include a second radius in order to compare the global and central star formation rate enhancements. Beginning with the solid lines in Figure \ref{fig:DMdot_DSFR_TPM_split}, we see that the two feedback modes show different behaviour in the first 500 Myrs post-merger. Radiative mode feedback galaxies show star formation rate enhancements for $\sim$ 500 Myrs after the merger. We therefore find that galaxies in radiative mode feedback, on average, have both accretion rate and star formation rate enhancements within the first few hundred Myrs. In contrast, kinetic mode feedback galaxies show a relative star formation \textit{suppression} for 500 Myrs after the merger. Therefore, Figure \ref{fig:DMdot_DSFR_TPM_split} demonstrates that despite an increased supply of gas, resulting in enhanced SMBH accretion rates, kinetic mode post-mergers display lower than average rates of star formation within $\sim $ 500 Myrs of coalescence. Our result that, on average, kinetic mode feedback galaxies demonstrate both relatively enhanced SMBH accretion rates and relatively suppressed of star formation rates may be explained considering the relationship between star formation quenching and kinetic mode feedback in TNG \citep[][]{Davies2020,Luo2020,Terrazas2020,Nelson2021,Piotrowska2021}, which we will discuss in further detail in Section \ref{subsec:feedback}.

Another possible explanation for the star formation rate suppression as well as the difference in timescales between $\Delta \dot M_{BH}$ and $\Delta SFR$ is that the enhancement in accretion rate is measured in the accretion region for the simulation, i.e. within the cells in the immediate vicinity of the black hole particle, whereas the star formation rate is measured over a much larger spatial scale, within twice the stellar half-mass radius. We therefore investigate whether central SFR enhancements are present in the PM sample. In the bottom panel of Figure \ref{fig:DMdot_DSFR_TPM_split}, the dashed lines represent the star formation rate enhancement calculated within one stellar half-mass radius, $\Delta SFR$\textsubscript{half}, once again split by feedback mode into the blue and red lines. Comparing the solid lines, $\Delta SFR$, with the dashed lines, $\Delta SFR$\textsubscript{half}, we can see how a more centralized aperture, although still much larger in spatial extent than the BH accretion region, affects the timescales of the star formation rate enhancements. For the radiative feedback mode galaxies, we see that the SFR enhancement peak is higher, reaching almost 0.4 dex. We also see that the enhancement is slightly longer lived. In the kinetic mode galaxies, we see that the star formation rate is slightly enhanced, 0.1 dex, within the $\sim$500 Myr window, demonstrating relative star formation rate enhancements despite suppressed star formation rates on a larger spatial scale.

\subsection{Correlation of star formation rate and accretion rate enhancements}
\label{subsec:correlations}

In the previous section, Figure \ref{fig:DMdot_DSFR_TPM} demonstrated that within $\sim $ 500 Myrs of the merger, on average, galaxies have both star formation rate enhancements and accretion rate enhancements. However, Figure \ref{fig:DMdot_DSFR_TPM} does not assess whether the two enhancements are temporally correlated in a given galaxy. Although the merger process triggers gas inflows \citep[][]{Hernquist1989a,Barnes1991,Mihos1996,DiMatteo2007,Capelo2016,Blumenthal2018} that might naturally lead to quasi-simultaneous enhancements in both nuclear star formation and BH accretion \citep[][]{Sanders1988,DiMatteo2005,Springel2005,Hopkins2008}, observations show little evidence for a starburst-AGN connection outside of the most luminous systems \citep[][]{RowanRobinson1995,Schweitzer2006,Lutz2010,Shao2010,Santini2012,Rosario2015}. High resolution binary merger simulations have also demonstrated the lack of temporal correlations for the majority of the duration of the merger, with \cite[][]{Volonteri20151} finding a correlation only during the 200-300 Myrs over which the galaxies coalesce. The challenge is the different timescales of the two processes; BH accretion is stochastic on very rapid timescales, whereas star formation is more sustained. Observations that take a snapshot of a single point in time can not capture any potential extended connection between these processes. Simulation snapshots suffer from the same effect. Nonetheless, a merger event pinpoints the time of major gas inflow and may therefore be expected to demonstrate a connection between star formation and accretion.

In order to test for a correlation between the SFR enhancement and SMBH accretion rate enhancements in post-mergr galaxies, and determine whether selecting the post-mergers within different windows of time post-merger may affect a measured correlation, we consider the relationship between $\Delta \dot M_{BH}$ and $\Delta SFR$ on a galaxy by galaxy basis. We note that for the remainder of this section, where we compare $\Delta \dot M_{BH}$ and $\Delta SFR$ for each individual galaxy, we use a single matching criteria to calculate both $\Delta \dot M_{BH}$ and $\Delta SFR$, where we include the star forming vs passive classification discussed in Section \ref{subsec:timescale} to our fiducial matching scheme outline in Section \ref{subsec:matching}. That is, for each post-merger we identify a set of suitable controls and calculate $\Delta \dot M_{BH}$ and $\Delta SFR$ relative to the same set of controls.

\begin{figure*}
	\includegraphics[width=2\columnwidth]{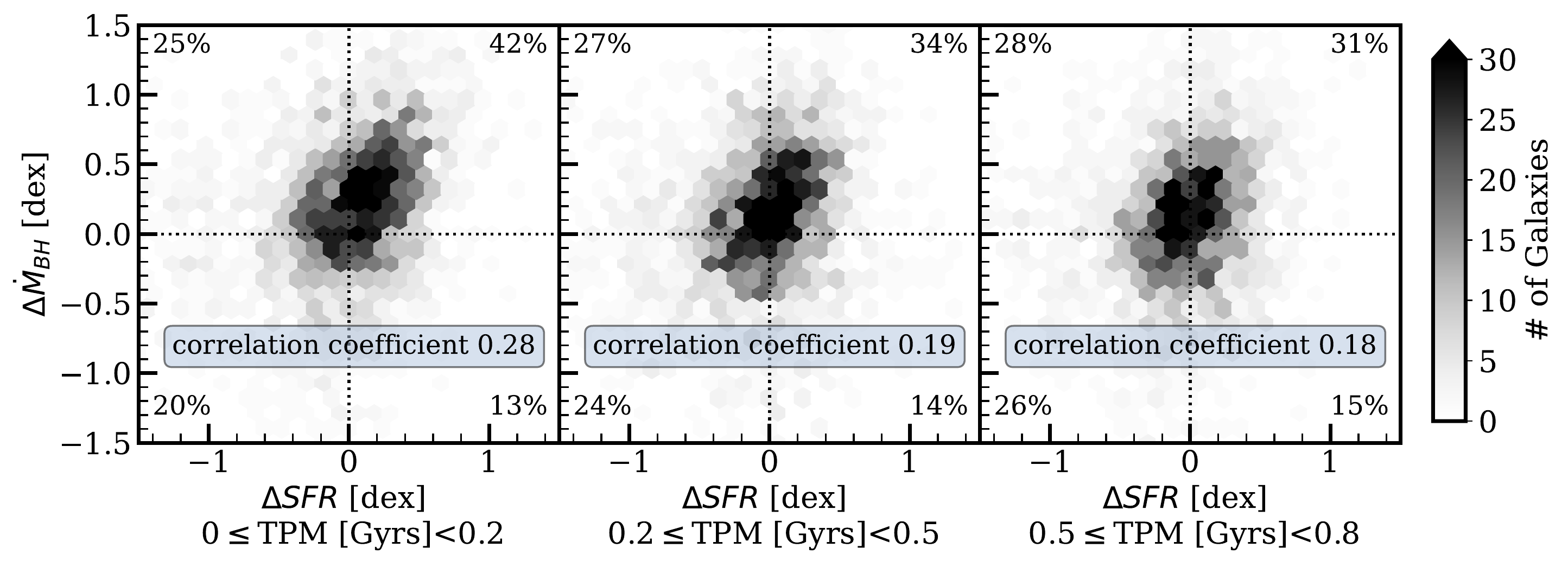}
     \caption{Accretion rate enhancements vs star formation rate enhancements, where each panel represents bins of time post-merger (TPM in Gyrs), since the most recent 1:10 mass ratio merger. The percentages represent the number of galaxies within each quadrant relative to the total number of galaxies in the panel. The Pearson correlation coefficient is shown in the light blue text box.}
    \label{fig:DMdot_vs_DSFR}
\end{figure*}

Figure \ref{fig:DMdot_vs_DSFR} shows the distribution of $\Delta \dot M_{BH}$ vs $\Delta SFR$ on a galaxy by galaxy basis, divided into three samples based on time post-merger. The percentages of galaxies occupying each quadrant are shown, where a high occupation fraction in the top right quadrant would be indicative of an excess of simultaneous enhancements.  We find that within 200 Myrs of the merger, 42\% of post-mergers have both accretion rate and SFR enhancements. However, the majority of post-mergers occupy each of the other quadrants, demonstrating that the star formation and accretion rate processes are not generally synchronized in post-mergers. We perform a Pearson correlation test for each window of time post-merger, with the correlation coefficient quoted in the textbox of each panel. The correlation coefficient is strongest within 200 Myrs of the merger and decreases with each subsequent time bin. However, even in the shortest time-since-merger interval the correlation between SFR and SMBH accretion rate enhancements is modest (correlation coefficient = 0.29). Our result therefore demonstrates that the synchronicity or correlation between SFR and SMBH accretion rate enhancements is affected by the timescale on which the galaxy is observed. We also demonstrate that a significant correlation between $\Delta \dot M_{BH}$ and $\Delta SFR$ is only present in the first few hundred Myrs post-merger, consistent with the timescale of temporally correlated SFR and SMBH accretion rates in \cite{Volonteri20151}. Finally, that even within 200 Myrs of coalescence, the majority of post-mergers do not have synchronized $\Delta SFR$ and $\Delta \dot M_{BH}$.

\begin{figure*}
	\includegraphics[width=1.5\columnwidth]{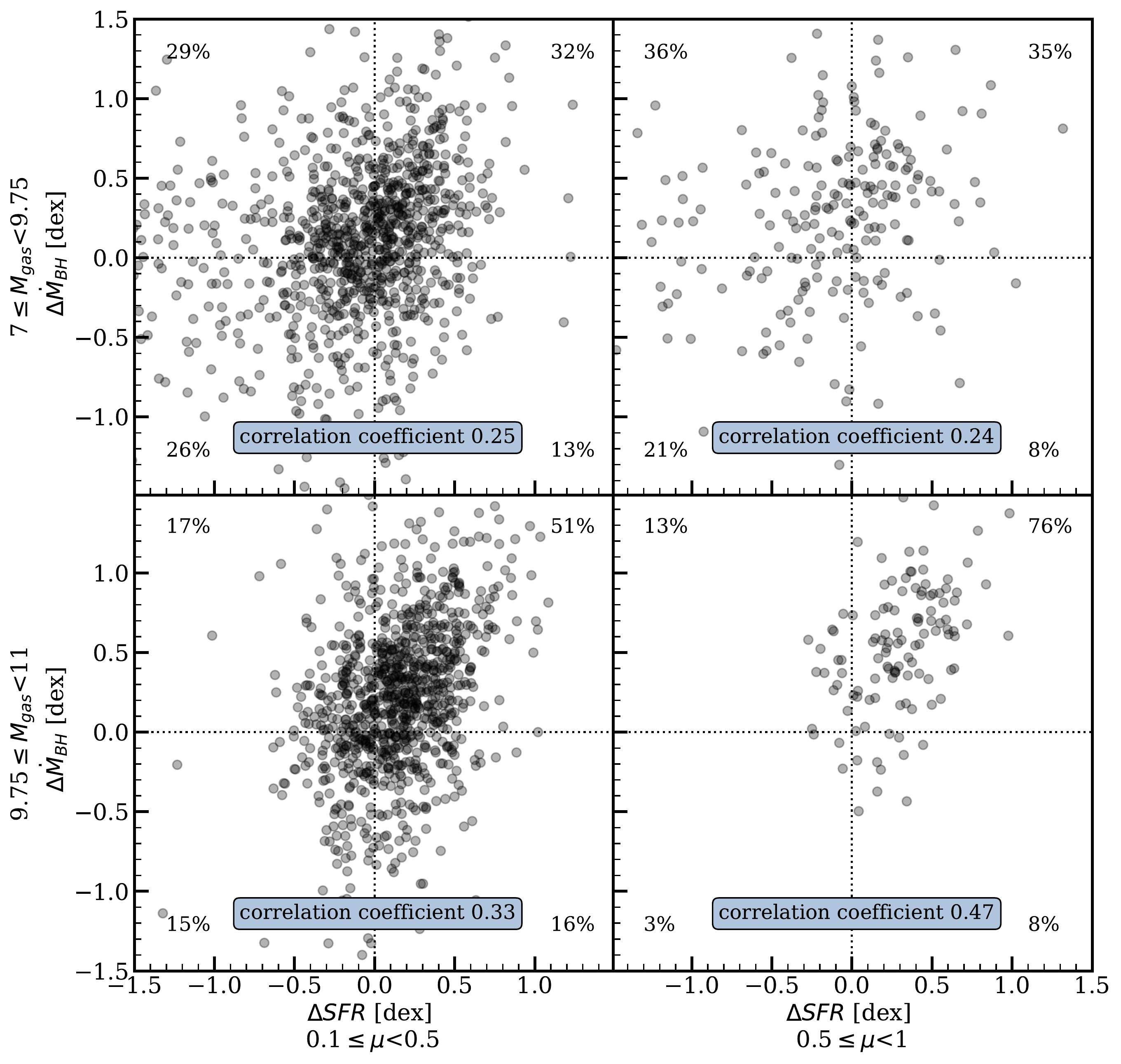}
    \caption{Accretion rate enhancement vs star formation rate enhancement for galaxies within 200 Myrs of a 1:10 mass ratio merger. The plots are organized by increasing gas mass, in units $\mathrm{log}_{10}\mathrm{M_{\odot}}$, from bottom to top and increasing mass ratio from left to right. The percentages represent the number of galaxies within each quadrant relative to the total number of galaxies in each panel. The Pearson correlation coefficient is shown in the light blue text box for each quadrant}
    \label{fig:DMdot_vs_DSFR_3x3}
\end{figure*}

Returning to the left-most panel of Figure \ref{fig:DMdot_vs_DSFR}, we note that despite a bias for co-incident positive $\Delta \dot M_{BH}$ and $\Delta SFR$, the majority of post-mergers do not demonstrate synchronicity in SFR and SMBH accretion rate enhancements. We explore the diversity of the post-mergers within 200 Myrs of the merger in Figure \ref{fig:DMdot_vs_DSFR_3x3}, which shows the distribution of the post-mergers in $\Delta \dot M_{BH}$ and $\Delta SFR$ space, separated into bins of mass ratio and gas mass. The panels are organized such that gas mass is increasing from top to bottom and mass ratio is increasing from left to right.

The post-mergers in the bottom row of panels of Figure \ref{fig:DMdot_vs_DSFR_3x3}, corresponding to a gas mass $10^{9.75-11} M_{\odot}$, demonstrate that the majority of gas-rich post-mergers have both positive $\Delta \dot M_{BH}$ and $\Delta SFR$ (i.e. over 50\% are in the top right quadrant). In fact, comparing the bottom left panel (gas-rich `minor' mergers) to the bottom right panel (gas-rich major mergers), we see that gas-rich major mergers are more likely to have synchronicity (51\% compared with 76\%). We also find that overall, gas rich major mergers are more likely to produce SMBH accretion rate enhancements (68\% positive $\Delta \dot M_{BH}$ compared with 89\%). Our result suggests that gas rich major mergers more reliably produce accretion rate enhancements compared with minor mergers, though not necessarily stronger enhancements (result of Figure \ref{fig:DMdot_vs_MassRatio}).

The panels of the top row of Figure \ref{fig:DMdot_vs_DSFR_3x3}, corresponding to gas mass $10^{7-9.75} M_{\odot}$, show a slight bias for galaxies to occupy the left-side quadrants, corresponding to relatively lower star formation rates. A star formation rate suppression in lower gas mass post-mergers is consistent with the results of Sections \ref{subsec:Enhancements} and \ref{subsec:timescale}. We also emphasize that we are considering the gas mass of the galaxy post-coalescence, an important distinction if comparing our results with previous work such as \cite{DiMatteo2007} and \cite{Scudder2015}. Specifically, \cite{DiMatteo2007} do not find a strong dependence of star formation rate enhancement on the initial gas mass (i.e. the amount of gas available just before coalescence), while \cite{Scudder2015} find that galaxies with the lowest initial gas fraction have the highest SFR enhancements due to interactions. Our results demonstrate that star formation rate enhancements are less likely in post-mergers with a low post-coalescence gas mass, but do not comment on the relationship with initial gas mass. Our dependence of $\Delta SFR$ on gas mass may be explained, in part, by the correlation between SFR and gas fraction (as was demonstrated in \citealt[][]{Scudder2015} and \citealt[][]{Hani2020}). However, we note that \cite{Hani2020} still find a dependence of $\Delta SFR$ on gas fraction even when explicitly control matching on gas mass. If we focus on the top right panel of Figure \ref{fig:DMdot_vs_DSFR_3x3}, we also find that the bias for negative $\Delta SFR$ is present even in the highest mass ratio mergers. Therefore, our results suggest that even major mergers are unlikely to produce star formation rate enhancements if they have a low gas mass.

Overall, Figure \ref{fig:DMdot_vs_DSFR_3x3} demonstrates that the strongest bias for synchronicity and the strongest correlation (coefficient=0.47) occurs in gas-rich major mergers. We note that for the results presented in Figure \ref{fig:DMdot_vs_DSFR_3x3}, we find the same qualitative and similar quantitative results when separating the post-mergers by gas fraction rather than gas mass.

\section{Discussion}
\label{sec:Discussion}

\subsection{Effect of Resolution and Numerical Considerations}
\label{subsec:ResolutionAndEtc}

\begin{figure}
	\includegraphics[width=\columnwidth]{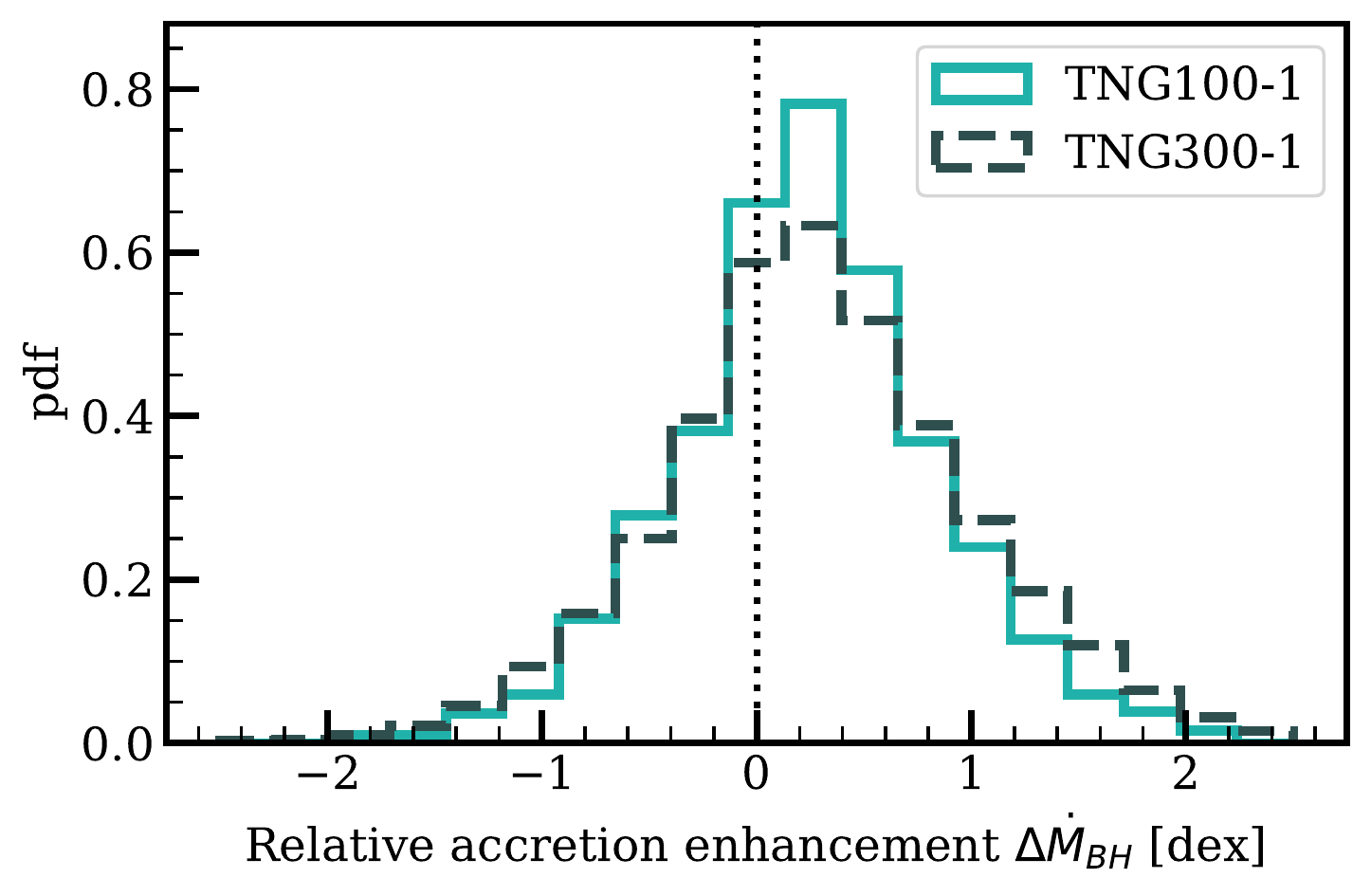}
    \caption{Histogram of accretion rate enhancements for the post-merger populations from the intermediate resolution, TNG100-1, and low resolution, TNG300, simulations. TNG100 is represented with the teal line, and TNG300 is represented with the dashed line.}
    \label{fig:DMdot_hist_res}
\end{figure}

In the work presented here, we use the intermediate volume and resolution run of TNG, TNG100-1. We can repeat our experiment using the large volume and low resolution run, TNG300-1, which has a $(302.6 \, \mathrm{Mpc})^3$ volume, a baryonic resolution of $1.1\times 10^7 \mathrm{M_{\odot}}$, and a dark matter resolution of $5.9 \times 10^7 \mathrm{M_{\odot}}$ \citep[][]{Weinberger2017,Pillepich2018}. We obtain a sample of 25576 successfully matched post-merger galaxies from TNG300-1. Figure \ref{fig:DMdot_hist_res} shows the accretion rate enhancement parameter for the post-merger sample from both resolutions, TNG100 shown in teal and TNG300 shown in the dashed line. The median $\Delta \dot M_{BH}$ of the TNG300 post-merger population is $\sim$0.27 dex, corresponding to an accretion rate enhancement of roughly 90\%, slightly higher but consistent with the median enhancement for TNG100. Therefore, we demonstrate the comparability of our results between the two resolutions and volumes. 

A limitation of any work looking at SMBH accretion rates using cosmological scale simulations is the subgrid model required to calculate $\dot M_{BH}$. Subgrid models are an essential part of cosmological simulations, as they include important physical processes, such as stellar and SMBH feedback, that occur on scales below the resolution of the simulation. As mentioned, the TNG physics model uses a Bondi-Hoyle-Lyttleton accretion model \citep[][]{Weinberger2017,Pillepich2018}. However, the accretion model includes simplifying assumptions such as spherical symmetry, which would not hold for an accretion disk of material surrounding a SMBH. \cite{NegriVolonteri2017} demonstrated the variability of accretion rates calculated by different Bondi models and the dependance on simulation choices such as how the local density and sound speed are sampled. They find that Bondi models often overestimate the accretion rate, particularly when AGN feedback is inefficient at evacuating gas from the central region of the galaxy. However, they also find that in some simulations the accretion rate can be underestimated, which occurs when simulation cells of hot gas are over-represented in calculation of local parameters. Therefore, while we are unable to comment on the effects of mergers on the subparsec scales around the SMBH, the work presented here provides insight on how a merger event can impact the supply of material to the nuclear region, which may go on to truly accrete onto the SMBH.

An additional limitation of this work concerns the implementation of SMBH relocation in cosmological simulations, and the subsequent effect on SMBH accretion and SMBH mergers. \cite{Bahe2022} investigate the effects of SMBH repositioning in simulations using the EAGLE galaxy physics model, and demonstrate that the repositioning of SMBHs to the potential minima can result in significant boosts to SMBH accretion rate due to the increased density at the gravitational potential minima and the reduced relative velocity of the SMBH after repositioning. In addition, SMBH repositioning promotes early SMBH merging as it places the SMBHs of the merging galaxies in close proximity. A higher SMBH mass, due to ‘premature’ merging, will promote higher accretion rates due to the dependence on SMBH mass in the Bondi model. Therefore, we may expect the TNG model to overestimate accretion rates in the PM sample since SMBHs in merging galaxies do not wander and may merge prematurely. In addition, SMBH mergers may further complicate the timescale and delay between processes like starbursts and AGN activity. For example, \cite{Ni2022} demonstrate using the ASTRID cosmological simulations, which do not anchor SMBH particles to the gravitational minima and instead include a subgrid recipe to compensate for unresolved dynamical friction, that there is a delay of $\sim $200 Myrs between first close encounter of SMBH pairs and the SMBH merger. The above limitations highlight the imperative to investigate the galaxy merger and AGN connection in other cosmological simulations with varying subgrid implementations for SMBH physics.

\subsection{Implications for feedback and quenching}
\label{subsec:feedback}

In Section \ref{subsec:timescale}, we suggested that the enhancement of SMBH accretion rates in post-mergers with kinetic mode AGN feedback may have implications for galaxy quenching in TNG. Numerous studies have demonstrated that in TNG effective kinetic-mode AGN feedback is responsible for galaxy quenching \citep[][]{Davies2020,Luo2020,Terrazas2020,Nelson2021,Piotrowska2021}. Here, we briefly comment on how merger induced accretion enhancements may influence galaxy quenching in TNG. For simplicity, we can consider idealistic scenarios through which the enhanced accretion rates in post-merger galaxies may affect the quenching process. 

First, increased accretion rates could result in post-merger galaxies that are in a high accretion state and are therefore using the radiative mode feedback model, which is ineffective at quenching in the TNG model. Therefore, mergers in TNG could inhibit or delay the quenching process, as galaxies may be in an inefficient feedback mode (in terms of quenching star formation). Since the median enhancement of accretion rates is subtle, $\sim1.7$ times higher than controls, a delayed quenching effect would be most applicable in galaxies that have accretion rates close to the transition between high and low accretion states, as defined in Section \ref{subsec:Illustris}. 

Alternatively, increased SMBH accretion rates may enhance quenching in TNG. First, accretion rate enhancements in galaxies with low accretion rates could increase the energy input via effective kinetic mode feedback and contribute to the quenching of galaxies. In addition, enhancement in accretion rate can promote growth of the SMBH. According to the TNG model, the transition between radiative and kinetic mode feedback is strongly dependent on the SMBH mass, such that high mass SMBHs are more likely to be have kinetic mode feedback. Therefore, post-mergers may transition towards effective kinetic mode feedback sooner, promoting an excess of quenching.

\cite{Quai2021} explicitly investigate quenching in post-merger galaxies in TNG and conclude that quenching is rare in post-merger galaxies, but that there is an excess of quenched post-mergers when compared with matched controls. Specifically, \cite{Quai2021} find that quenching in post-mergers occurs only in galaxies that were already predisposed to quenching. The rarity of quenched post-mergers would be consistent with the first scenario outlined above. Post-mergers with strongly enhanced accretion rates will likely be in radiative feedback mode, where the SMBH feedback would not interact with the host galaxy in a way that promotes quenching, leading to a rarity of quenched post-merger galaxies. Furthermore, the merger process may introduce more gas into the nuclear region of the host galaxy, prolonging the lifetime of the high accretion state and preventing quenching. 

The second scenario we have proposed is consistent with the slight excess of quenched post-mergers found in \cite{Quai2021}, where increased effective kinetic mode feedback speeds up quenching in the post-merger galaxies that are predisposed to quench. Furthermore, a truncation of star formation in the massive post-merger galaxies following a period of enhanced AGN activity would be consistent with the findings of \cite{Dubois2016} who observe a decrease of star formation in massive galaxies due to merger induced AGN activity. In addition, we discuss how our results relate to observational studies in Section \ref{subsec:observations}.

\subsection{What fraction of mergers are AGN?}
\label{subsec:fractionAGN}

So far we have demonstrated that post-merger galaxies have (on average) enhanced accretion rates. We have also demonstrated that not all post-mergers show enhancements in their SMBH accretion rate. Next we address the question of how frequently mergers actually trigger an AGN. Specifically, in this experiment, we will quantify what fraction of all post-mergers will have a high accretion event within 500 Myrs of coalescence.

\begin{figure}
	\includegraphics[width=\columnwidth]{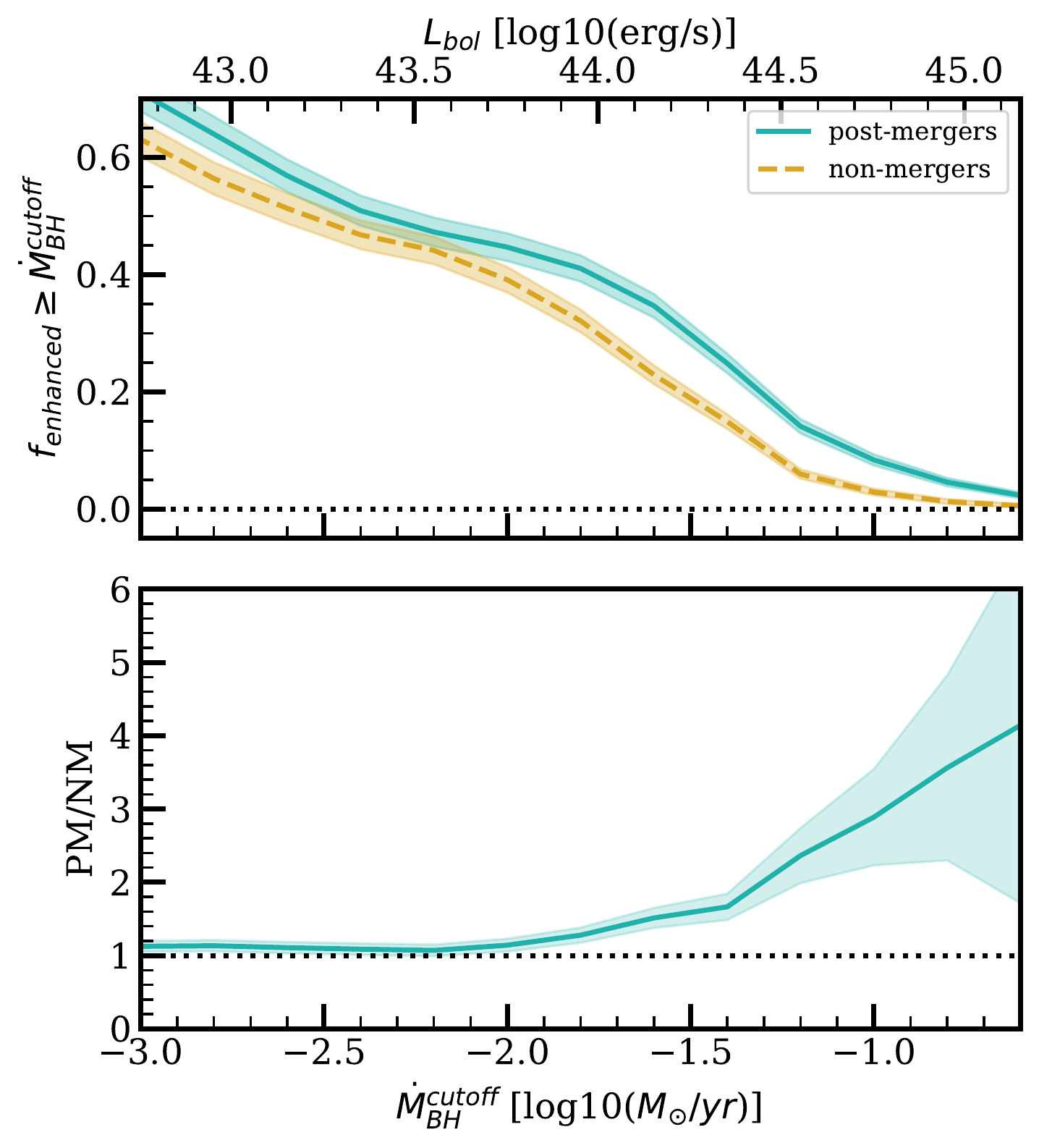}
    \caption{The top panel shows the fraction of galaxies that have a SMBH accretion rate at or higher than $\dot M_{BH}^{\mathrm{cutoff}}$ within a period of 500 Myrs, relative to the total number of galaxies. The teal line corresponds to the post-merger sample and the yellow dashed line is the non-merger sample, as defined in Section \ref{subsec: PM and NM}. The top x-axis corresponds to the AGN luminosity, calculated as $L_{\mathrm{bol}}=0.1\dot M_{BH}c^2$. The bottom panel shows the ratio of the fraction of post-merger galaxies to non-merger galaxies, or the fractional excess of AGN in the post-merger sample. The error in the shaded region is the Poisson error reflecting the number of galaxies.}
    \label{fig:fEnhancedAcc}
\end{figure}

To begin, we define an enhancement fraction, $f_{\mathrm{enhanced}}$, as the number of galaxies that have an SMBH accretion rate at or above a cutoff, $\dot M_{BH}^{\mathrm{cutoff}}$, divided by the total number of galaxies in the sample. We calculate $f_{\mathrm{enhanced}}$ for both the post-merger and non-merger sample, defined in Section \ref{subsec: PM and NM}, following the post-merger for 500 Myrs after the merger and the non-merger sample for 500 Myrs of secular evolution. The result of the experiment is shown in Figure \ref{fig:fEnhancedAcc}.

The top panel of Figure \ref{fig:fEnhancedAcc} shows $f_{\mathrm{enhanced}}$ as a function of $\dot M_{BH}^{\mathrm{cutoff}}$, where the post-merger sample is shown in the solid teal line and the non-merger sample is shown in the dashed yellow line. The top axis shows the associated bolometric luminosity for the SMBH accretion rate cutoff, calculated as 10\% of the accretion mass energy, or $0.1 \dot M_{BH}c^2$. We demonstrate that $\sim$ 60\% of post-merger galaxies have accretion rates exceeding $L_{bol} > 10^{43} \mathrm{erg/s}$ within 500 Myrs of the merger. However, we find that 50\% of non-merger galaxies also have an AGN phase of $L_{bol} > 10^{43} \mathrm{erg/s}$ within the same time period. Our result demonstrates that accretion rate events exceeding $L_{bol} > 10^{43} \mathrm{erg/s}$ are common in both samples, though slightly more common in the post-mergers, as expected.

In the bottom panel of Figure \ref{fig:fEnhancedAcc} we show the ratio of the post-merger to non-merger $f_{\mathrm{enhanced}}$ (the ratio of the blue solid line to the yellow dashed line from the top panel), or the fractional excess of AGN in the post-merger. We demonstrate that even though the fraction of galaxies that have an AGN phase decreases with increasing luminosity for both samples (as expected), the fractional excess of post-mergers with an AGN phase increases with luminosity. Therefore while less than 10\% of post-mergers will achieve accretion rates exceeding $L_{bol} > 10^{45} \mathrm{erg/s}$, four times more AGN appear in the post-merger sample than the non-merger sample. 

Overall, we find that only a small fraction of mergers have high accretion rate events but that post-mergers are more likely to have an AGN event than a non-merger as a function of luminosity, consistent with results from the Magneticum Pathfinder simulation \citep[][]{Steinborn2018} and the EAGLE simulation \citep[][]{McAlpine2020}, as well as complimentary to \cite{Bhowmick2020} who find that high luminosity AGN are more likely to appear in environments with a higher density of SMBHs in TNG.

\subsection{What fraction of AGN are mergers?}
\label{subsec:fractionMergers}

A complementary question to the one posed in Section \ref{subsec:fractionAGN} is what fraction of galaxies with high accretion rates are mergers? In Section \ref{subsec:fractionAGN}, we demonstrated that the post-merger sample galaxies are more likely to be AGN than the non-merger sample of galaxies. However, post-mergers are rare in both TNG and the observed universe and thus may not be the major pathway to AGN triggering. Therefore, we measure the merger fraction of AGN in TNG100-1 in order to quantify the contribution of mergers to the total AGN population in the simulation.

\begin{figure}
	\includegraphics[width=\columnwidth]{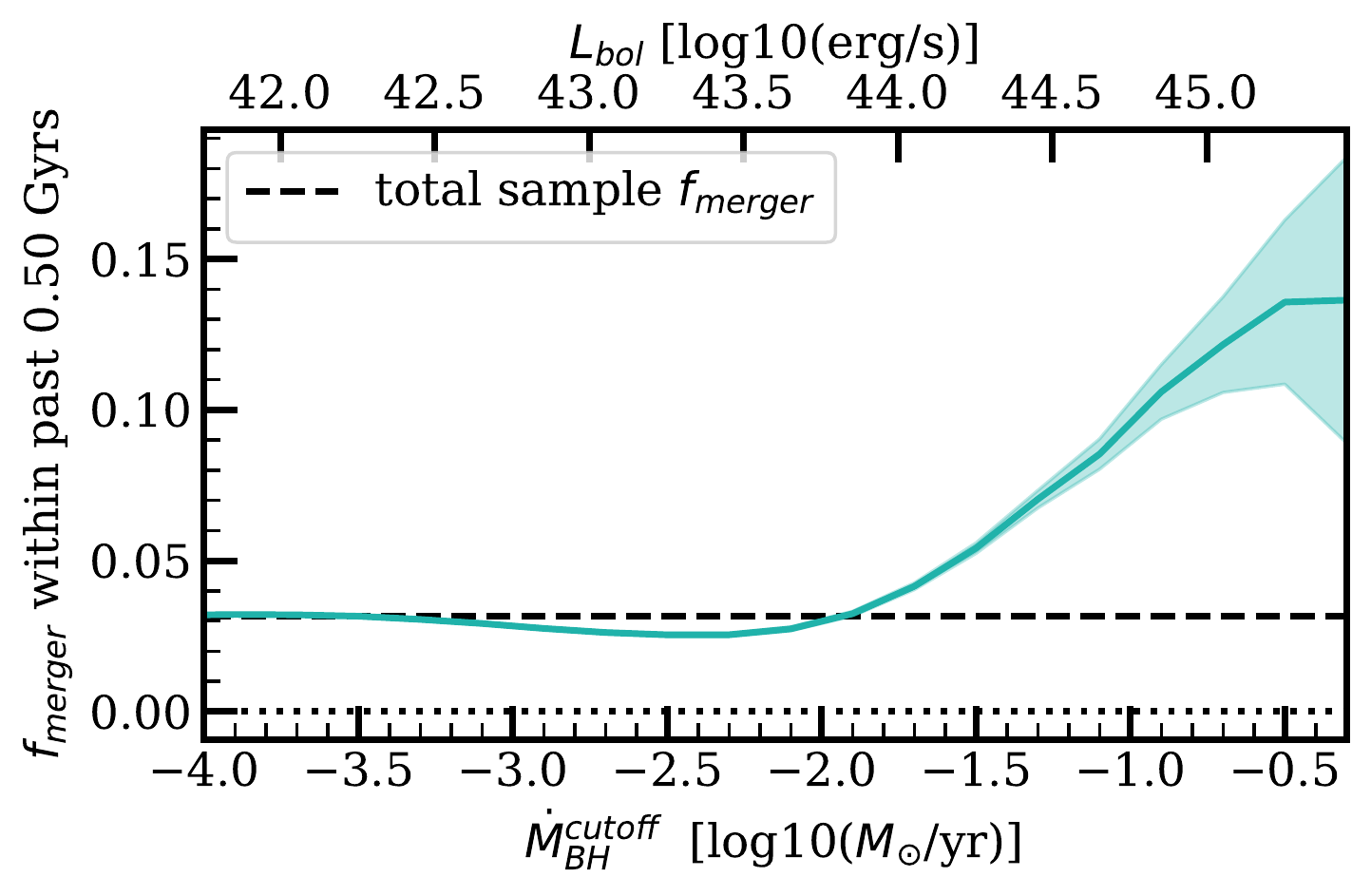}
    \caption{The teal line shows the fraction of galaxies that have had a merger of mass ratio greater than 1:10 within 500 the past Myrs. The x-axis defines the sample used to calculate the merger fraction from left to right, where only galaxies with a black hole accretion rate of at least $\dot M_{BH}^{\mathrm{cutoff}}$ are included in the calculation of the merger fraction. The top x-axis shows the corresponding AGN luminosity, calculated as $L_{\mathrm{bol}}=0.1\dot M_{BH}c^2$. The error in the shaded region is the Poisson error reflecting the number of galaxies. The horizontal dashed line represents the overall merger fraction of the sample, $\sim$3\%, for all galaxies in TNG with stellar mass of at least $10^{10} M_{\odot}$ and a redshift $< 1$.}
    \label{fig:fMerger}
\end{figure}

To investigate the merger fraction of AGN, we first select an `AGN sample' by selecting all galaxies that have a SMBH accretion rate at or above $\dot M_{BH}^{\mathrm{cutoff}}$. In this experiment, the merger fraction is defined as the fraction of galaxies that have had a merger in the last 500 Myrs. Figure \ref{fig:fMerger} shows the merger fraction as a function of the cutoff accretion rate. The x-axis defines the AGN sample used to calculate the merger fraction, where the AGN sample will consist of all galaxies with accretion rates at or exceeding $\dot M_{BH}^{\mathrm{cutoff}}$ (and meeting the selection criteria outlined in Section \ref{subsec: ID mergers}, i.e $M_{\star} > 10^{10} \mathrm{M_{\odot}}$ and $z<1$). Once again, we show the equivalent bolometric luminosity along the top axis, calculated as $L_{bol} = 0.1 \dot M_{BH}c^2$. Figure \ref{fig:fMerger} shows that the merger fraction increases as a function of accretion rate (or, equivalently, luminosity). The horizontal dashed line represents the total merger fraction for the entire sample of galaxies, $f_{\mathrm{merger}}\sim$ 3\%. We see that the merger fraction of the AGN sample exceeds the total $f_{\mathrm{merger}}$ around $L_{bol}\sim 10^{44} \mathrm{erg/s}$. Therefore, there is no significant luminosity dependence of the merger fraction for $L_{bol}\lesssim 10^{44} \mathrm{erg/s}$. Beyond $L_{bol}\sim 10^{44} \mathrm{erg/s}$ the merger fraction increases with AGN luminosity, with a peak merger fraction of $\sim$13\%.

Although there is a luminosity dependence on the merger fraction for $L_{bol}\gtrsim 10^{44} \mathrm{erg/s}$, we find that recent mergers never dominate the AGN sample, even at the highest luminosities. This result is once again consistent with the results of \cite{Steinborn2018} and \cite{McAlpine2020} for the Magneticum Pathfinder and EAGLE simulations. We note that the merger fraction that we have defined does not include galaxies in the pre-merger phase, which may contribute to the remaining $\sim$90\% of galaxies at or above $10^{45}$ erg/s. There may also be minor mergers that are not accounted for in the merger fraction. However, the result of Figure \ref{fig:fMerger} also supports a scenario where secular processes play a significant role in fueling the highest luminosity AGN (at least in the redshift regime 0-1). This experiment concludes that post-mergers contribute more to the AGN population at the highest accretion rates, but they are non-dominant over the entire accretion rate range.

\subsection{Comparisons to observations}
\label{subsec:observations}
In the previous sections, we have demonstrated that there is a population averaged enhancement in SMBH accretion rate in post-mergers and that post-mergers are more likely to host highly accreting SMBHs. Our findings are in agreement with the observational studies that find that AGN are more likely to appear in a post-merger or interacting galaxy sample when compared with undisturbed controls \citep{Woods2007,Alonso2007,Koss2010,Ellison2011,RamosAlmeida2012,Ellison2013,Satyapal2014,Hong2015,Kocevski2015,Rosario2015,Weston2017,Hewlett2017,Goulding2018,Ellison2019,Gao2020,Marian2020,Pierce2022}. We find that at most, AGN occur four times more commonly in the post-mergers compared with the non-mergers. The maximum excess of AGN is quantitatively consistent with numerous observational studies finding an excess of approximately four or lower \citep[][]{Ellison2013,Goulding2018,Gao2020}, though some studies find a significantly larger excess of AGN in merging galaxies \citep[][]{Satyapal2014,Weston2017}. Such variations in the observed excess may be, in part, due to the different AGN selection techniques, which have been demonstrated to result in different AGN fractions \citep[][]{Secrest2020,Bickley2022Submitted}. Variations may also exist within the same AGN selection techniques, as demonstrated in \cite{Bickley2022Submitted} who find the AGN excess may also depend on the strength of visual disturbances in the PM sample..

In addition, we demonstrate that the majority of high luminosity AGN are not associated to recent mergers. Our results are in agreement with observations which find the most luminous AGN are not mergers \citep{Cisternas2011,Schawinski2011,Schawinski2012,Kocevski2012,Bohm2013,Villforth2014,Mechtley2016,Hewlett2017,Villforth2017,Marian2019,Lambrides2021}, but are in contrast to observational studies that find the majority of quasars have disturbed morphology \citep[][]{Treister2012,Glikman2015,Fan2016,Goulding2018,UrbanoMayorgas2019}. Therefore, our results support a scenario where mergers can, but do not always, trigger AGN, and where mergers do not play a major role is triggering even the most luminous AGN. 

Despite an increased likelihood to host highly accreting SMBHs, we demonstrate that the majority of mergers do not host a highly accreting SMBH, suggesting the majority of mergers will not undergo a strong AGN feedback event. In fact, observations have demonstrated that mergers have normal (or even enhanced) gas fractions (e.g. \citealt{Ellison20152,Ellison2018,Violino2018,Pan2018}), consistent with a lack of strong instantaneous feedback. However, observations do find that AGN can have effects on the small (kpc or below) scale interstellar medium (e.g. \citealt[][]{Oosterloo2017,Izumi2020,Ellison2021,GarciaBurillo2021,RamosAlmeida2022,Saito2022}).

In Section \ref{subsec:correlations}, we demonstrated that the strength of the star formation rate and SMBH accretion rate enhancements, on a galaxy by galaxy basis, are not generally correlated, in agreement with simulations such as \cite{Hickox2014,Volonteri20151}. However, we do find a connection between star formation rate and SMBH accretion rate in post-mergers with both a high mass ratio and significant amount of gas. Such a connection may reflect the general preference for AGN to reside in gas rich and star forming galaxies, as demonstrated in simulations \citep[][]{Ward2022} and observations \citep[][]{Rosario20131,Bernhard2016,Jarvis2020,Ellison20192,Shangguan2020,Xie2021,Koss2021}. In addition, we demonstrate in Figure \ref{fig:DMdot_vs_DSFR_3x3} numerous individual cases where a post-merger may have a strongly enhanced SMBH accretion rate and simultaneously suppressed or normal star formation rate relative to controls. Our results are therefore consistent with observations which show that most AGN (not strictly post-mergers) have typical rates of star formation \citep[][]{Rosario20131,Rosario2015}, although some observations demonstrate higher than normal star formation rates in the highest luminosity AGN \citep[][]{Schweitzer2006,Lutz2010,Shao2010,Santini2012}.

Finally, in Section \ref{subsec:ResolutionAndEtc}, we comment the consistency between our results and the rarity of rapid quenching in TNG post-mergers found in \cite{Quai2021}. However, \cite{Ellison2022} demonstrate that in observations, there exists a significant excess of rapidly/recently quenched post starburst galaxies (PSBs) in galaxy mergers. There is still uncertainty about the quenching mechanisms within PSBs, as observations show they do not lack molecular gas (e.g. \citealt{Rowlands2015,French2015}). Instead, it might be that PSBs lack dense star-forming gas \citep[][]{French20181}, perhaps due to enhanced turbulence (e.g. \citealt[][]{Smercina2022}). Such mechanisms may not be captured within TNG due to the resolution constraints and effective equation of state treatment of the interstellar medium, warranting an investigation of AGN and rapid quenching in post-merger galaxies using higher resolution simulations or simulations with varied physical models.

\section{Conclusions}
\label{sec:conclusion}

In the work presented here, we study the effect of galaxy mergers on SMBH accretion rates in a collection of 1563 post-merger galaxies from the IllustrisTNG simulation. Our post-merger sample reflects a diverse collection of stellar masses, star formation rates, gas masses, environments, and mass ratios. For each post-merger galaxy in our sample, we identify control galaxies for comparison, and are able to isolate the effect of the merger on the instantaneous SMBH accretion rates. Our results are summarized in the following points:
\begin{itemize}
    \item On average, post-merger black hole accretion rates are enhanced by a factor of 0.23 dex, corresponding to a $\sim$70\% increase in accretion rate (Figure \ref{fig:DMdot_hist_all}). We find that there is significant variation in the accretion rate enhancement on a galaxy by galaxy basis, where $\sim$ 30\% of post-mergers have accretion rates consistent with no enhancement or lower than controls.
    
    \medskip
    
    \item SMBH accretion rate enhancements persist for up to two Gyrs after coalescence, and are significantly longer lived than SFR enhancements, which only persist for $\sim$500 Myrs post-merger (Figure \ref{fig:DMdot_DSFR_TPM}). 
    
    \medskip
    
    \item We find that the co-incidence of accretion rate and SFR enhancements is most pronounced within the first few hundred Myrs post-merger, and that the correlation strength of the enhancements decreases with time post-merger, shown in Figure \ref{fig:DMdot_vs_DSFR}. However, even within 200 Myrs of coalescence, the majority of post-mergers do not demonstrate synchronicity in SMBH acretion rate enhancements and SFR enhancements. 
    
    \medskip
    
    \item In Figure \ref{fig:DMdot_vs_DSFR_3x3}, we find that gas rich major mergers demonstrate the strongest correlation coefficient between $\Delta SFR$ and $\Delta \dot M_{BH}$, and the majority have enhancements in both. However, we find that the presence of gas alone is an insufficient criteria for enhancing SMBH accretion rates, and that gas rich non-mergers, on average, do not have positive $\Delta \dot M_{BH}$ (Figure \ref{fig:DMdot_vs_Mgas}).
    
    \medskip
    
    \item Post-mergers are more likely to host high luminosity AGN than non-merger galaxies (Figure \ref{fig:fEnhancedAcc}), however high luminosity AGN are rare in the post-merger sample. For AGN luminosity in excess of $L_{\mathrm{bol}} > 10^{45} \mathrm{erg/s}$, corresponding to $\sim$10\% of post-mergers, there are four times more AGN in the post-merger sample than the matched non-merger sample. 
    
    \medskip
    
    \item We find a luminosity dependence in the merger fraction of AGN for $L_{\mathrm{bol}} > 10^{44} \mathrm{erg/s}$ (Figure \ref{fig:fMerger}). However, we find that the majority of high luminosity AGN are not recent mergers, which at most only contribute $\sim$13\% to AGN with $L_{\mathrm{bol}} > 10^{45} \mathrm{erg/s}$.
\end{itemize}

\section*{Acknowledgements}
We thank the IllustrisTNG collaboration for making their data accessible. We thank the anonymous referee for their constructive comments. SBM acknowledges the receipt of a British Columbia Graduate Scholarship and the Dr. Margaret Perkins Hess Research Fellowship from the University of Victoria. This research was enabled in part by support provided by WestGrid (www.westgrid.ca) and Compute Canada (www.computecanada.ca). DRP gratefully acknowledges NSERC of Canada for a Discovery Grant which helped to fund this research.

\section*{Data Availability}
The data used in this work are publicly available at \hyperlink{}{https://www.tng-project.org}.



\bibliographystyle{mnras}
\bibliography{paperBib}

\begin{thebibliography}{}
\makeatletter
\relax
\def\mn@urlcharsother{\let\do\@makeother \do\$\do\&\do\#\do\^\do\_\do\%\do\~}
\def\mn@doi{\begingroup\mn@urlcharsother \@ifnextchar [ {\mn@doi@}
  {\mn@doi@[]}}
\def\mn@doi@[#1]#2{\def\@tempa{#1}\ifx\@tempa\@empty \href
  {http://dx.doi.org/#2} {doi:#2}\else \href {http://dx.doi.org/#2} {#1}\fi
  \endgroup}
\def\mn@eprint#1#2{\mn@eprint@#1:#2::\@nil}
\def\mn@eprint@arXiv#1{\href {http://arxiv.org/abs/#1} {{\tt arXiv:#1}}}
\def\mn@eprint@dblp#1{\href {http://dblp.uni-trier.de/rec/bibtex/#1.xml}
  {dblp:#1}}
\def\mn@eprint@#1:#2:#3:#4\@nil{\def\@tempa {#1}\def\@tempb {#2}\def\@tempc
  {#3}\ifx \@tempc \@empty \let \@tempc \@tempb \let \@tempb \@tempa \fi \ifx
  \@tempb \@empty \def\@tempb {arXiv}\fi \@ifundefined
  {mn@eprint@\@tempb}{\@tempb:\@tempc}{\expandafter \expandafter \csname
  mn@eprint@\@tempb\endcsname \expandafter{\@tempc}}}

\bibitem[\protect\citeauthoryear{{Alonso}, {Lambas}, {Tissera}  \&
  {Coldwell}}{{Alonso} et~al.}{2007}]{Alonso2007}
{Alonso} M.~S.,  {Lambas} D.~G.,  {Tissera} P.,   {Coldwell} G.,  2007, \mn@doi
  [\mnras] {10.1111/j.1365-2966.2007.11367.x}, \href
  {https://ui.adsabs.harvard.edu/abs/2007MNRAS.375.1017A} {375, 1017}

\bibitem[\protect\citeauthoryear{{Bah{\'e}} et~al.,}{{Bah{\'e}}
  et~al.}{2022}]{Bahe2022}
{Bah{\'e}} Y.~M.,  et~al., 2022, \mn@doi [\mnras] {10.1093/mnras/stac1339},
  \href {https://ui.adsabs.harvard.edu/abs/2022MNRAS.516..167B} {516, 167}

\bibitem[\protect\citeauthoryear{{Barnes} \& {Hernquist}}{{Barnes} \&
  {Hernquist}}{1991}]{Barnes1991}
{Barnes} J.~E.,  {Hernquist} L.~E.,  1991, \mn@doi [\apjl] {10.1086/185978},
  \href {https://ui.adsabs.harvard.edu/abs/1991ApJ...370L..65B} {370, L65}

\bibitem[\protect\citeauthoryear{{Barton}, {Geller}  \& {Kenyon}}{{Barton}
  et~al.}{2000}]{Barton2000}
{Barton} E.~J.,  {Geller} M.~J.,   {Kenyon} S.~J.,  2000, \mn@doi [\apj]
  {10.1086/308392}, \href
  {https://ui.adsabs.harvard.edu/abs/2000ApJ...530..660B} {530, 660}

\bibitem[\protect\citeauthoryear{{Bernhard}, {Mullaney}, {Daddi}, {Ciesla}  \&
  {Schreiber}}{{Bernhard} et~al.}{2016}]{Bernhard2016}
{Bernhard} E.,  {Mullaney} J.~R.,  {Daddi} E.,  {Ciesla} L.,   {Schreiber} C.,
  2016, \mn@doi [\mnras] {10.1093/mnras/stw973}, \href
  {https://ui.adsabs.harvard.edu/abs/2016MNRAS.460..902B} {460, 902}

\bibitem[\protect\citeauthoryear{{Bernhard}, {Tadhunter}, {Pierce}, {Dicken},
  {Mullaney}, {Morganti}, {Ramos Almeida}  \& {Daddi}}{{Bernhard}
  et~al.}{2022}]{Bernhard2022}
{Bernhard} E.,  {Tadhunter} C.~N.,  {Pierce} J.~C.~S.,  {Dicken} D.,
  {Mullaney} J.~R.,  {Morganti} R.,  {Ramos Almeida} C.,   {Daddi} E.,  2022,
  \mn@doi [\mnras] {10.1093/mnras/stac474}, \href
  {https://ui.adsabs.harvard.edu/abs/2022MNRAS.tmp..489B} {}

\bibitem[\protect\citeauthoryear{{Bessiere}, {Tadhunter}, {Ramos Almeida}  \&
  {Villar Mart{\'\i}n}}{{Bessiere} et~al.}{2012}]{Bessiere2012}
{Bessiere} P.~S.,  {Tadhunter} C.~N.,  {Ramos Almeida} C.,   {Villar
  Mart{\'\i}n} M.,  2012, \mn@doi [\mnras] {10.1111/j.1365-2966.2012.21701.x},
  \href {https://ui.adsabs.harvard.edu/abs/2012MNRAS.426..276B} {426, 276}

\bibitem[\protect\citeauthoryear{{Bhowmick}, {Blecha}  \& {Thomas}}{{Bhowmick}
  et~al.}{2020}]{Bhowmick2020}
{Bhowmick} A.~K.,  {Blecha} L.,   {Thomas} J.,  2020, \mn@doi [\apj]
  {10.3847/1538-4357/abc1e6}, \href
  {https://ui.adsabs.harvard.edu/abs/2020ApJ...904..150B} {904, 150}

\bibitem[\protect\citeauthoryear{{Bickley}, {Ellison}, R.  \&
  {Wilkinson}}{{Bickley} et~al.}{2022b}]{Bickley2022Submitted}
{Bickley} R.~W.,  {Ellison} S.~L.,  R. P.~D.,   {Wilkinson} S.,  2022b,
  submitted to \mnras

\bibitem[\protect\citeauthoryear{{Bickley}, {Ellison}, {Patton}, {Bottrell},
  {Gwyn}  \& {Hudson}}{{Bickley} et~al.}{2022a}]{Bickley2022}
{Bickley} R.~W.,  {Ellison} S.~L.,  {Patton} D.~R.,  {Bottrell} C.,  {Gwyn} S.,
    {Hudson} M.~J.,  2022a, \mn@doi [\mnras] {10.1093/mnras/stac1500}, \href
  {https://ui.adsabs.harvard.edu/abs/2022MNRAS.tmp.1474B} {}

\bibitem[\protect\citeauthoryear{Blumenthal \& Barnes}{Blumenthal \&
  Barnes}{2018}]{Blumenthal2018}
Blumenthal K.~A.,  Barnes J.~E.,  2018, \mn@doi [\mnras]
  {10.1093/mnras/sty1605}, 479, 3952

\bibitem[\protect\citeauthoryear{{Blumenthal} et~al.,}{{Blumenthal}
  et~al.}{2020}]{Blumenthal2020}
{Blumenthal} K.~A.,  et~al., 2020, \mn@doi [\mnras] {10.1093/mnras/stz3472},
  \href {https://ui.adsabs.harvard.edu/abs/2020MNRAS.492.2075B} {492, 2075}

\bibitem[\protect\citeauthoryear{{B{\"o}hm} et~al.,}{{B{\"o}hm}
  et~al.}{2013}]{Bohm2013}
{B{\"o}hm} A.,  et~al., 2013, \mn@doi [\aap] {10.1051/0004-6361/201015444},
  \href {https://ui.adsabs.harvard.edu/abs/2013A&A...549A..46B} {549, A46}

\bibitem[\protect\citeauthoryear{{Cao} et~al.,}{{Cao} et~al.}{2016}]{Cao2016}
{Cao} C.,  et~al., 2016, \mn@doi [\apjs] {10.3847/0067-0049/222/2/16}, \href
  {https://ui.adsabs.harvard.edu/abs/2016ApJS..222...16C} {222, 16}

\bibitem[\protect\citeauthoryear{Capelo \& Dotti}{Capelo \&
  Dotti}{2016}]{Capelo2016}
Capelo P.~R.,  Dotti M.,  2016, \mn@doi [\mnras] {10.1093/mnras/stw2872}, 465,
  2643

\bibitem[\protect\citeauthoryear{{Capelo}, {Volonteri}, {Dotti}, {Bellovary},
  {Mayer}  \& {Governato}}{{Capelo} et~al.}{2015}]{Capelo2015}
{Capelo} P.~R.,  {Volonteri} M.,  {Dotti} M.,  {Bellovary} J.~M.,  {Mayer} L.,
   {Governato} F.,  2015, \mn@doi [\mnras] {10.1093/mnras/stu2500}, \href
  {https://ui.adsabs.harvard.edu/abs/2015MNRAS.447.2123C} {447, 2123}

\bibitem[\protect\citeauthoryear{{Casteels} et~al.,}{{Casteels}
  et~al.}{2014}]{Casteels2014}
{Casteels} K. R.~V.,  et~al., 2014, \mn@doi [\mnras] {10.1093/mnras/stu1799},
  \href {https://ui.adsabs.harvard.edu/abs/2014MNRAS.445.1157C} {445, 1157}

\bibitem[\protect\citeauthoryear{{Chiaberge}, {Gilli}, {Lotz}  \&
  {Norman}}{{Chiaberge} et~al.}{2015}]{Chiaberge2015}
{Chiaberge} M.,  {Gilli} R.,  {Lotz} J.~M.,   {Norman} C.,  2015, \mn@doi
  [\apj] {10.1088/0004-637X/806/2/147}, \href
  {https://ui.adsabs.harvard.edu/abs/2015ApJ...806..147C} {806, 147}

\bibitem[\protect\citeauthoryear{{Cisternas} et~al.,}{{Cisternas}
  et~al.}{2011}]{Cisternas2011}
{Cisternas} M.,  et~al., 2011, \mn@doi [\apj] {10.1088/0004-637X/726/2/57},
  \href {https://ui.adsabs.harvard.edu/abs/2011ApJ...726...57C} {726, 57}

\bibitem[\protect\citeauthoryear{{Davies}, {Crain}, {Oppenheimer}  \&
  {Schaye}}{{Davies} et~al.}{2020}]{Davies2020}
{Davies} J.~J.,  {Crain} R.~A.,  {Oppenheimer} B.~D.,   {Schaye} J.,  2020,
  \mn@doi [\mnras] {10.1093/mnras/stz3201}, \href
  {https://ui.adsabs.harvard.edu/abs/2020MNRAS.491.4462D} {491, 4462}

\bibitem[\protect\citeauthoryear{{Di Matteo}, {Springel}  \& {Hernquist}}{{Di
  Matteo} et~al.}{2005}]{DiMatteo2005}
{Di Matteo} T.,  {Springel} V.,   {Hernquist} L.,  2005, \mn@doi [\nat]
  {10.1038/nature03335}, \href
  {https://ui.adsabs.harvard.edu/abs/2005Natur.433..604D} {433, 604}

\bibitem[\protect\citeauthoryear{{Di Matteo}, {Combes}, {Melchior}  \&
  {Semelin}}{{Di Matteo} et~al.}{2007}]{DiMatteo2007}
{Di Matteo} P.,  {Combes} F.,  {Melchior} A.~L.,   {Semelin} B.,  2007, \mn@doi
  [\aap] {10.1051/0004-6361:20066959}, \href
  {https://ui.adsabs.harvard.edu/abs/2007A&A...468...61D} {468, 61}

\bibitem[\protect\citeauthoryear{{Di Matteo}, {Bournaud}, {Martig}, {Combes},
  {Melchior}  \& {Semelin}}{{Di Matteo} et~al.}{2008}]{DiMatteo2008}
{Di Matteo} P.,  {Bournaud} F.,  {Martig} M.,  {Combes} F.,  {Melchior} A.~L.,
   {Semelin} B.,  2008, \mn@doi [\aap] {10.1051/0004-6361:200809480}, \href
  {https://ui.adsabs.harvard.edu/abs/2008A&A...492...31D} {492, 31}

\bibitem[\protect\citeauthoryear{{Dicken} et~al.,}{{Dicken}
  et~al.}{2012}]{Dicken2012}
{Dicken} D.,  et~al., 2012, \mn@doi [\apj] {10.1088/0004-637X/745/2/172}, \href
  {https://ui.adsabs.harvard.edu/abs/2012ApJ...745..172D} {745, 172}

\bibitem[\protect\citeauthoryear{{Donley} et~al.,}{{Donley}
  et~al.}{2018}]{Donley2018}
{Donley} J.~L.,  et~al., 2018, \mn@doi [\apj] {10.3847/1538-4357/aa9ffa}, \href
  {https://ui.adsabs.harvard.edu/abs/2018ApJ...853...63D} {853, 63}

\bibitem[\protect\citeauthoryear{{Donnari} et~al.,}{{Donnari}
  et~al.}{2019}]{Donnari2019}
{Donnari} M.,  et~al., 2019, \mn@doi [\mnras] {10.1093/mnras/stz712}, \href
  {https://ui.adsabs.harvard.edu/abs/2019MNRAS.485.4817D} {485, 4817}

\bibitem[\protect\citeauthoryear{{Dubois}, {Peirani}, {Pichon}, {Devriendt},
  {Gavazzi}, {Welker}  \& {Volonteri}}{{Dubois} et~al.}{2016}]{Dubois2016}
{Dubois} Y.,  {Peirani} S.,  {Pichon} C.,  {Devriendt} J.,  {Gavazzi} R.,
  {Welker} C.,   {Volonteri} M.,  2016, \mn@doi [\mnras]
  {10.1093/mnras/stw2265}, \href
  {https://ui.adsabs.harvard.edu/abs/2016MNRAS.463.3948D} {463, 3948}

\bibitem[\protect\citeauthoryear{{Dunn}, {Fender}, {K{\"o}rding}, {Belloni}  \&
  {Cabanac}}{{Dunn} et~al.}{2010}]{Dunn2010}
{Dunn} R.~J.~H.,  {Fender} R.~P.,  {K{\"o}rding} E.~G.,  {Belloni} T.,
  {Cabanac} C.,  2010, \mn@doi [\mnras] {10.1111/j.1365-2966.2010.16114.x},
  \href {https://ui.adsabs.harvard.edu/abs/2010MNRAS.403...61D} {403, 61}

\bibitem[\protect\citeauthoryear{Ellison, Patton, Simard  \&
  McConnachie}{Ellison et~al.}{2008}]{Ellison2008}
Ellison S.~L.,  Patton D.~R.,  Simard L.,   McConnachie A.~W.,  2008, \mn@doi
  [\aj] {10.1088/0004-6256/135/5/1877}, 135, 1877

\bibitem[\protect\citeauthoryear{{Ellison}, {Patton}, {Mendel}  \&
  {Scudder}}{{Ellison} et~al.}{2011}]{Ellison2011}
{Ellison} S.~L.,  {Patton} D.~R.,  {Mendel} J.~T.,   {Scudder} J.~M.,  2011,
  \mn@doi [\mnras] {10.1111/j.1365-2966.2011.19624.x}, \href
  {https://ui.adsabs.harvard.edu/abs/2011MNRAS.418.2043E} {418, 2043}

\bibitem[\protect\citeauthoryear{{Ellison}, {Mendel}, {Patton}  \&
  {Scudder}}{{Ellison} et~al.}{2013}]{Ellison2013}
{Ellison} S.~L.,  {Mendel} J.~T.,  {Patton} D.~R.,   {Scudder} J.~M.,  2013,
  \mn@doi [\mnras] {10.1093/mnras/stt1562}, \href
  {https://ui.adsabs.harvard.edu/abs/2013MNRAS.435.3627E} {435, 3627}

\bibitem[\protect\citeauthoryear{{Ellison}, {Fertig}, {Rosenberg}, {Nair},
  {Simard}, {Torrey}  \& {Patton}}{{Ellison} et~al.}{2015a}]{Ellison20152}
{Ellison} S.~L.,  {Fertig} D.,  {Rosenberg} J.~L.,  {Nair} P.,  {Simard} L.,
  {Torrey} P.,   {Patton} D.~R.,  2015a, \mn@doi [\mnras]
  {10.1093/mnras/stu2744}, \href
  {https://ui.adsabs.harvard.edu/abs/2015MNRAS.448..221E} {448, 221}

\bibitem[\protect\citeauthoryear{{Ellison}, {Patton}  \& {Hickox}}{{Ellison}
  et~al.}{2015b}]{Ellison2015}
{Ellison} S.~L.,  {Patton} D.~R.,   {Hickox} R.~C.,  2015b, \mn@doi [\mnras]
  {10.1093/mnrasl/slv061}, \href
  {https://ui.adsabs.harvard.edu/abs/2015MNRAS.451L..35E} {451, L35}

\bibitem[\protect\citeauthoryear{{Ellison}, {Catinella}  \&
  {Cortese}}{{Ellison} et~al.}{2018}]{Ellison2018}
{Ellison} S.~L.,  {Catinella} B.,   {Cortese} L.,  2018, \mn@doi [\mnras]
  {10.1093/mnras/sty1247}, \href
  {https://ui.adsabs.harvard.edu/abs/2018MNRAS.478.3447E} {478, 3447}

\bibitem[\protect\citeauthoryear{{Ellison}, {Brown}, {Catinella}  \&
  {Cortese}}{{Ellison} et~al.}{2019a}]{Ellison20192}
{Ellison} S.~L.,  {Brown} T.,  {Catinella} B.,   {Cortese} L.,  2019a, \mn@doi
  [\mnras] {10.1093/mnras/sty3139}, \href
  {https://ui.adsabs.harvard.edu/abs/2019MNRAS.482.5694E} {482, 5694}

\bibitem[\protect\citeauthoryear{{Ellison}, {Viswanathan}, {Patton},
  {Bottrell}, {McConnachie}, {Gwyn}  \& {Cuillandre}}{{Ellison}
  et~al.}{2019b}]{Ellison2019}
{Ellison} S.~L.,  {Viswanathan} A.,  {Patton} D.~R.,  {Bottrell} C.,
  {McConnachie} A.~W.,  {Gwyn} S.,   {Cuillandre} J.-C.,  2019b, \mn@doi
  [\mnras] {10.1093/mnras/stz1431}, \href
  {https://ui.adsabs.harvard.edu/abs/2019MNRAS.487.2491E} {487, 2491}

\bibitem[\protect\citeauthoryear{{Ellison} et~al.,}{{Ellison}
  et~al.}{2021}]{Ellison2021}
{Ellison} S.~L.,  et~al., 2021, \mn@doi [\mnras] {10.1093/mnrasl/slab047},
  \href {https://ui.adsabs.harvard.edu/abs/2021MNRAS.505L..46E} {505, L46}

\bibitem[\protect\citeauthoryear{{Ellison} et~al.,}{{Ellison}
  et~al.}{2022}]{Ellison2022}
{Ellison} S.~L.,  et~al., 2022, \mn@doi [\mnras] {10.1093/mnrasl/slac109},
  \href {https://ui.adsabs.harvard.edu/abs/2022MNRAS.517L..92E} {517, L92}

\bibitem[\protect\citeauthoryear{Fan et~al.,}{Fan et~al.}{2016}]{Fan2016}
Fan L.,  et~al., 2016, \mn@doi [\apj] {10.3847/2041-8205/822/2/l32}, 822, L32

\bibitem[\protect\citeauthoryear{{French}, {Yang}, {Zabludoff}, {Narayanan},
  {Shirley}, {Walter}, {Smith}  \& {Tremonti}}{{French}
  et~al.}{2015}]{French2015}
{French} K.~D.,  {Yang} Y.,  {Zabludoff} A.,  {Narayanan} D.,  {Shirley} Y.,
  {Walter} F.,  {Smith} J.-D.,   {Tremonti} C.~A.,  2015, \mn@doi [\apj]
  {10.1088/0004-637X/801/1/1}, \href
  {https://ui.adsabs.harvard.edu/abs/2015ApJ...801....1F} {801, 1}

\bibitem[\protect\citeauthoryear{{French}, {Zabludoff}, {Yoon}, {Shirley},
  {Yang}, {Smercina}, {Smith}  \& {Narayanan}}{{French}
  et~al.}{2018}]{French20181}
{French} K.~D.,  {Zabludoff} A.~I.,  {Yoon} I.,  {Shirley} Y.,  {Yang} Y.,
  {Smercina} A.,  {Smith} J.~D.,   {Narayanan} D.,  2018, \mn@doi [\apj]
  {10.3847/1538-4357/aac8de}, \href
  {https://ui.adsabs.harvard.edu/abs/2018ApJ...861..123F} {861, 123}

\bibitem[\protect\citeauthoryear{{Gao} et~al.,}{{Gao} et~al.}{2020}]{Gao2020}
{Gao} F.,  et~al., 2020, \mn@doi [\aap] {10.1051/0004-6361/201937178}, \href
  {https://ui.adsabs.harvard.edu/abs/2020A&A...637A..94G} {637, A94}

\bibitem[\protect\citeauthoryear{{Garc{\'\i}a-Burillo}
  et~al.,}{{Garc{\'\i}a-Burillo} et~al.}{2021}]{GarciaBurillo2021}
{Garc{\'\i}a-Burillo} S.,  et~al., 2021, \mn@doi [\aap]
  {10.1051/0004-6361/202141075}, \href
  {https://ui.adsabs.harvard.edu/abs/2021A&A...652A..98G} {652, A98}

\bibitem[\protect\citeauthoryear{{Glikman}, {Simmons}, {Mailly}, {Schawinski},
  {Urry}  \& {Lacy}}{{Glikman} et~al.}{2015}]{Glikman2015}
{Glikman} E.,  {Simmons} B.,  {Mailly} M.,  {Schawinski} K.,  {Urry} C.~M.,
  {Lacy} M.,  2015, \mn@doi [\apj] {10.1088/0004-637X/806/2/218}, \href
  {https://ui.adsabs.harvard.edu/abs/2015ApJ...806..218G} {806, 218}

\bibitem[\protect\citeauthoryear{{Goulding} et~al.,}{{Goulding}
  et~al.}{2018}]{Goulding2018}
{Goulding} A.~D.,  et~al., 2018, \mn@doi [\pasj] {10.1093/pasj/psx135}, \href
  {https://ui.adsabs.harvard.edu/abs/2018PASJ...70S..37G} {70, S37}

\bibitem[\protect\citeauthoryear{Hani, Gosain, Ellison, Patton  \& Torrey}{Hani
  et~al.}{2020}]{Hani2020}
Hani M.~H.,  Gosain H.,  Ellison S.~L.,  Patton D.~R.,   Torrey P.,  2020,
  \mn@doi [\mnras] {10.1093/mnras/staa459}, 493, 3716

\bibitem[\protect\citeauthoryear{{Hernquist}}{{Hernquist}}{1989}]{Hernquist1989a}
{Hernquist} L.,  1989, \mn@doi [\nat] {10.1038/340687a0}, \href
  {https://ui.adsabs.harvard.edu/abs/1989Natur.340..687H} {340, 687}

\bibitem[\protect\citeauthoryear{{Hewlett}, {Villforth}, {Wild},
  {Mendez-Abreu}, {Pawlik}  \& {Rowlands}}{{Hewlett}
  et~al.}{2017}]{Hewlett2017}
{Hewlett} T.,  {Villforth} C.,  {Wild} V.,  {Mendez-Abreu} J.,  {Pawlik} M.,
  {Rowlands} K.,  2017, \mn@doi [\mnras] {10.1093/mnras/stx997}, \href
  {https://ui.adsabs.harvard.edu/abs/2017MNRAS.470..755H} {470, 755}

\bibitem[\protect\citeauthoryear{{Hickox} \& {Alexander}}{{Hickox} \&
  {Alexander}}{2018}]{HickoxAlexander2018}
{Hickox} R.~C.,  {Alexander} D.~M.,  2018, \mn@doi [\araa]
  {10.1146/annurev-astro-081817-051803}, \href
  {https://ui.adsabs.harvard.edu/abs/2018ARA&A..56..625H} {56, 625}

\bibitem[\protect\citeauthoryear{{Hickox}, {Mullaney}, {Alexander}, {Chen},
  {Civano}, {Goulding}  \& {Hainline}}{{Hickox} et~al.}{2014}]{Hickox2014}
{Hickox} R.~C.,  {Mullaney} J.~R.,  {Alexander} D.~M.,  {Chen} C.-T.~J.,
  {Civano} F.~M.,  {Goulding} A.~D.,   {Hainline} K.~N.,  2014, \mn@doi [\apj]
  {10.1088/0004-637X/782/1/9}, \href
  {https://ui.adsabs.harvard.edu/abs/2014ApJ...782....9H} {782, 9}

\bibitem[\protect\citeauthoryear{Hong, Im, Kim  \& Ho}{Hong
  et~al.}{2015}]{Hong2015}
Hong J.,  Im M.,  Kim M.,   Ho L.~C.,  2015, \mn@doi [\apj]
  {10.1088/0004-637x/804/1/34}, 804, 34

\bibitem[\protect\citeauthoryear{{Hopkins}, {Hernquist}, {Cox}  \&
  {Kere{\v{s}}}}{{Hopkins} et~al.}{2008}]{Hopkins2008}
{Hopkins} P.~F.,  {Hernquist} L.,  {Cox} T.~J.,   {Kere{\v{s}}} D.,  2008,
  \mn@doi [\apjs] {10.1086/524362}, \href
  {https://ui.adsabs.harvard.edu/abs/2008ApJS..175..356H} {175, 356}

\bibitem[\protect\citeauthoryear{{Hopkins}, {Cox}, {Hernquist}, {Narayanan},
  {Hayward}  \& {Murray}}{{Hopkins} et~al.}{2013}]{Hopkins2013}
{Hopkins} P.~F.,  {Cox} T.~J.,  {Hernquist} L.,  {Narayanan} D.,  {Hayward}
  C.~C.,   {Murray} N.,  2013, \mn@doi [\mnras] {10.1093/mnras/stt017}, \href
  {https://ui.adsabs.harvard.edu/abs/2013MNRAS.430.1901H} {430, 1901}

\bibitem[\protect\citeauthoryear{{Izumi} et~al.,}{{Izumi}
  et~al.}{2020}]{Izumi2020}
{Izumi} T.,  et~al., 2020, \mn@doi [\apj] {10.3847/1538-4357/ab99a8}, \href
  {https://ui.adsabs.harvard.edu/abs/2020ApJ...898...61I} {898, 61}

\bibitem[\protect\citeauthoryear{{Jarvis} et~al.,}{{Jarvis}
  et~al.}{2020}]{Jarvis2020}
{Jarvis} M.~E.,  et~al., 2020, \mn@doi [\mnras] {10.1093/mnras/staa2196}, \href
  {https://ui.adsabs.harvard.edu/abs/2020MNRAS.498.1560J} {498, 1560}

\bibitem[\protect\citeauthoryear{{Kennicutt}}{{Kennicutt}}{1998}]{Kennicutt1998}
{Kennicutt} Robert~C. J.,  1998, \mn@doi [\apj] {10.1086/305588}, \href
  {https://ui.adsabs.harvard.edu/abs/1998ApJ...498..541K} {498, 541}

\bibitem[\protect\citeauthoryear{{Knapen}, {Cisternas}  \&
  {Querejeta}}{{Knapen} et~al.}{2015}]{Knapen2015}
{Knapen} J.~H.,  {Cisternas} M.,   {Querejeta} M.,  2015, \mn@doi [\mnras]
  {10.1093/mnras/stv2135}, \href
  {https://ui.adsabs.harvard.edu/abs/2015MNRAS.454.1742K} {454, 1742}

\bibitem[\protect\citeauthoryear{{Kocevski} et~al.,}{{Kocevski}
  et~al.}{2012}]{Kocevski2012}
{Kocevski} D.~D.,  et~al., 2012, \mn@doi [\apj] {10.1088/0004-637X/744/2/148},
  \href {https://ui.adsabs.harvard.edu/abs/2012ApJ...744..148K} {744, 148}

\bibitem[\protect\citeauthoryear{{Kocevski} et~al.,}{{Kocevski}
  et~al.}{2015}]{Kocevski2015}
{Kocevski} D.~D.,  et~al., 2015, \mn@doi [\apj] {10.1088/0004-637X/814/2/104},
  \href {https://ui.adsabs.harvard.edu/abs/2015ApJ...814..104K} {814, 104}

\bibitem[\protect\citeauthoryear{{Koss}, {Mushotzky}, {Veilleux}  \&
  {Winter}}{{Koss} et~al.}{2010}]{Koss2010}
{Koss} M.,  {Mushotzky} R.,  {Veilleux} S.,   {Winter} L.,  2010, \mn@doi
  [\apjl] {10.1088/2041-8205/716/2/L125}, \href
  {https://ui.adsabs.harvard.edu/abs/2010ApJ...716L.125K} {716, L125}

\bibitem[\protect\citeauthoryear{{Koss} et~al.,}{{Koss}
  et~al.}{2021}]{Koss2021}
{Koss} M.~J.,  et~al., 2021, \mn@doi [\apjs] {10.3847/1538-4365/abcbfe}, \href
  {https://ui.adsabs.harvard.edu/abs/2021ApJS..252...29K} {252, 29}

\bibitem[\protect\citeauthoryear{{Lambrides} et~al.,}{{Lambrides}
  et~al.}{2021}]{Lambrides2021}
{Lambrides} E.~L.,  et~al., 2021, \mn@doi [\apj] {10.3847/1538-4357/ac12c8},
  \href {https://ui.adsabs.harvard.edu/abs/2021ApJ...919..129L} {919, 129}

\bibitem[\protect\citeauthoryear{{Luo}, {Li}, {Kang}, {Li}  \& {Wang}}{{Luo}
  et~al.}{2020}]{Luo2020}
{Luo} Y.,  {Li} Z.,  {Kang} X.,  {Li} Z.,   {Wang} P.,  2020, \mn@doi [\mnras]
  {10.1093/mnrasl/slaa099}, \href
  {https://ui.adsabs.harvard.edu/abs/2020MNRAS.496L.116L} {496, L116}

\bibitem[\protect\citeauthoryear{{Lutz} et~al.,}{{Lutz}
  et~al.}{2010}]{Lutz2010}
{Lutz} D.,  et~al., 2010, \mn@doi [\apj] {10.1088/0004-637X/712/2/1287}, \href
  {https://ui.adsabs.harvard.edu/abs/2010ApJ...712.1287L} {712, 1287}

\bibitem[\protect\citeauthoryear{{Marian} et~al.,}{{Marian}
  et~al.}{2019}]{Marian2019}
{Marian} V.,  et~al., 2019, \mn@doi [\apj] {10.3847/1538-4357/ab385b}, \href
  {https://ui.adsabs.harvard.edu/abs/2019ApJ...882..141M} {882, 141}

\bibitem[\protect\citeauthoryear{{Marian} et~al.,}{{Marian}
  et~al.}{2020}]{Marian2020}
{Marian} V.,  et~al., 2020, \mn@doi [\apj] {10.3847/1538-4357/abbd3e}, \href
  {https://ui.adsabs.harvard.edu/abs/2020ApJ...904...79M} {904, 79}

\bibitem[\protect\citeauthoryear{Marinacci et~al.,}{Marinacci
  et~al.}{2018}]{Mariancci2018}
Marinacci F.,  et~al., 2018, \mn@doi [\mnras] {10.1093/mnras/sty2206}, 480,
  5113

\bibitem[\protect\citeauthoryear{{McAlpine}, {Harrison}, {Rosario},
  {Alexander}, {Ellison}, {Johansson}  \& {Patton}}{{McAlpine}
  et~al.}{2020}]{McAlpine2020}
{McAlpine} S.,  {Harrison} C.~M.,  {Rosario} D.~J.,  {Alexander} D.~M.,
  {Ellison} S.~L.,  {Johansson} P.~H.,   {Patton} D.~R.,  2020, \mn@doi
  [\mnras] {10.1093/mnras/staa1123}, \href
  {https://ui.adsabs.harvard.edu/abs/2020MNRAS.494.5713M} {494, 5713}

\bibitem[\protect\citeauthoryear{{Mechtley} et~al.,}{{Mechtley}
  et~al.}{2016}]{Mechtley2016}
{Mechtley} M.,  et~al., 2016, \mn@doi [\apj] {10.3847/0004-637X/830/2/156},
  \href {https://ui.adsabs.harvard.edu/abs/2016ApJ...830..156M} {830, 156}

\bibitem[\protect\citeauthoryear{{Mihos} \& {Hernquist}}{{Mihos} \&
  {Hernquist}}{1996}]{Mihos1996}
{Mihos} J.~C.,  {Hernquist} L.,  1996, \mn@doi [\apj] {10.1086/177353}, \href
  {https://ui.adsabs.harvard.edu/abs/1996ApJ...464..641M} {464, 641}

\bibitem[\protect\citeauthoryear{{Moreno}, {Torrey}, {Ellison}, {Patton},
  {Bluck}, {Bansal}  \& {Hernquist}}{{Moreno} et~al.}{2015}]{Moreno2015}
{Moreno} J.,  {Torrey} P.,  {Ellison} S.~L.,  {Patton} D.~R.,  {Bluck} A.
  F.~L.,  {Bansal} G.,   {Hernquist} L.,  2015, \mn@doi [\mnras]
  {10.1093/mnras/stv094}, \href
  {https://ui.adsabs.harvard.edu/abs/2015MNRAS.448.1107M} {448, 1107}

\bibitem[\protect\citeauthoryear{{Moreno} et~al.,}{{Moreno}
  et~al.}{2019}]{Moreno2019}
{Moreno} J.,  et~al., 2019, \mn@doi [\mnras] {10.1093/mnras/stz417}, \href
  {https://ui.adsabs.harvard.edu/abs/2019MNRAS.485.1320M} {485, 1320}

\bibitem[\protect\citeauthoryear{Naiman et~al.,}{Naiman
  et~al.}{2018}]{Naiman2018}
Naiman J.~P.,  et~al., 2018, \mn@doi [\mnras] {10.1093/mnras/sty618}, 477, 1206

\bibitem[\protect\citeauthoryear{Negri \& Volonteri}{Negri \&
  Volonteri}{2017}]{NegriVolonteri2017}
Negri A.,  Volonteri M.,  2017, \mn@doi [\mnras] {10.1093/mnras/stx362}, 467,
  3475

\bibitem[\protect\citeauthoryear{Nelson et~al.,}{Nelson
  et~al.}{2017}]{Nelson2017}
Nelson D.,  et~al., 2017, \mn@doi [\mnras] {10.1093/mnras/stx3040}, 475, 624

\bibitem[\protect\citeauthoryear{{Nelson} et~al.,}{{Nelson}
  et~al.}{2021}]{Nelson2021}
{Nelson} E.~J.,  et~al., 2021, \mn@doi [\mnras] {10.1093/mnras/stab2131}, \href
  {https://ui.adsabs.harvard.edu/abs/2021MNRAS.508..219N} {508, 219}

\bibitem[\protect\citeauthoryear{{Nevin}, {Blecha}, {Comerford}  \&
  {Greene}}{{Nevin} et~al.}{2019}]{Nevin2019}
{Nevin} R.,  {Blecha} L.,  {Comerford} J.,   {Greene} J.,  2019, \mn@doi [\apj]
  {10.3847/1538-4357/aafd34}, \href
  {https://ui.adsabs.harvard.edu/abs/2019ApJ...872...76N} {872, 76}

\bibitem[\protect\citeauthoryear{{Ni} et~al.,}{{Ni} et~al.}{2022}]{Ni2022}
{Ni} Y.,  et~al., 2022, \mn@doi [\mnras] {10.1093/mnras/stac351}, \href
  {https://ui.adsabs.harvard.edu/abs/2022MNRAS.513..670N} {513, 670}

\bibitem[\protect\citeauthoryear{{Oosterloo}, {Raymond Oonk}, {Morganti},
  {Combes}, {Dasyra}, {Salom{\'e}}, {Vlahakis}  \& {Tadhunter}}{{Oosterloo}
  et~al.}{2017}]{Oosterloo2017}
{Oosterloo} T.,  {Raymond Oonk} J.~B.,  {Morganti} R.,  {Combes} F.,  {Dasyra}
  K.,  {Salom{\'e}} P.,  {Vlahakis} N.,   {Tadhunter} C.,  2017, \mn@doi [\aap]
  {10.1051/0004-6361/201731781}, \href
  {https://ui.adsabs.harvard.edu/abs/2017A&A...608A..38O} {608, A38}

\bibitem[\protect\citeauthoryear{{Pan} et~al.,}{{Pan} et~al.}{2018}]{Pan2018}
{Pan} H.-A.,  et~al., 2018, \mn@doi [\apj] {10.3847/1538-4357/aaeb92}, \href
  {https://ui.adsabs.harvard.edu/abs/2018ApJ...868..132P} {868, 132}

\bibitem[\protect\citeauthoryear{{Patton}, {Torrey}, {Ellison}, {Mendel}  \&
  {Scudder}}{{Patton} et~al.}{2013}]{Patton2013}
{Patton} D.~R.,  {Torrey} P.,  {Ellison} S.~L.,  {Mendel} J.~T.,   {Scudder}
  J.~M.,  2013, \mn@doi [\mnras] {10.1093/mnrasl/slt058}, \href
  {https://ui.adsabs.harvard.edu/abs/2013MNRAS.433L..59P} {433, L59}

\bibitem[\protect\citeauthoryear{{Patton}, {Qamar}, {Ellison}, {Bluck},
  {Simard}, {Mendel}, {Moreno}  \& {Torrey}}{{Patton}
  et~al.}{2016}]{Patton2016}
{Patton} D.~R.,  {Qamar} F.~D.,  {Ellison} S.~L.,  {Bluck} A. F.~L.,  {Simard}
  L.,  {Mendel} J.~T.,  {Moreno} J.,   {Torrey} P.,  2016, \mn@doi [\mnras]
  {10.1093/mnras/stw1494}, \href
  {https://ui.adsabs.harvard.edu/abs/2016MNRAS.461.2589P} {461, 2589}

\bibitem[\protect\citeauthoryear{Patton et~al.,}{Patton
  et~al.}{2020}]{Patton2020}
Patton D.~R.,  et~al., 2020, \mn@doi [\mnras] {10.1093/mnras/staa913}, 494,
  4969

\bibitem[\protect\citeauthoryear{{Pierce} et~al.,}{{Pierce}
  et~al.}{2022}]{Pierce2022}
{Pierce} J.~C.~S.,  et~al., 2022, \mn@doi [\mnras] {10.1093/mnras/stab3231},
  \href {https://ui.adsabs.harvard.edu/abs/2022MNRAS.510.1163P} {510, 1163}

\bibitem[\protect\citeauthoryear{Pillepich et~al.,}{Pillepich
  et~al.}{2017}]{Pillepich20171}
Pillepich A.,  et~al., 2017, \mn@doi [\mnras] {10.1093/mnras/stx3112}, 475, 648

\bibitem[\protect\citeauthoryear{{Pillepich} et~al.,}{{Pillepich}
  et~al.}{2018}]{Pillepich2018}
{Pillepich} A.,  et~al., 2018, \mn@doi [\mnras] {10.1093/mnras/stx2656}, \href
  {https://ui.adsabs.harvard.edu/abs/2018MNRAS.473.4077P} {473, 4077}

\bibitem[\protect\citeauthoryear{{Piotrowska}, {Bluck}, {Maiolino}  \&
  {Peng}}{{Piotrowska} et~al.}{2021}]{Piotrowska2021}
{Piotrowska} J.~M.,  {Bluck} A. F.~L.,  {Maiolino} R.,   {Peng} Y.,  2021,
  \mn@doi [\mnras] {10.1093/mnras/stab3673}, \href
  {https://ui.adsabs.harvard.edu/abs/2021MNRAS.tmp.3414P} {}

\bibitem[\protect\citeauthoryear{Quai, Hani, Ellison, Patton  \& Woo}{Quai
  et~al.}{2021}]{Quai2021}
Quai S.,  Hani M.~H.,  Ellison S.~L.,  Patton D.~R.,   Woo J.,  2021, \mn@doi
  [\mnras] {10.1093/mnras/stab988}, 504, 1888

\bibitem[\protect\citeauthoryear{{Ramos Almeida}, {Tadhunter}, {Inskip},
  {Morganti}, {Holt}  \& {Dicken}}{{Ramos Almeida}
  et~al.}{2011}]{RamosAlmeida2011}
{Ramos Almeida} C.,  {Tadhunter} C.~N.,  {Inskip} K.~J.,  {Morganti} R.,
  {Holt} J.,   {Dicken} D.,  2011, \mn@doi [\mnras]
  {10.1111/j.1365-2966.2010.17542.x}, \href
  {https://ui.adsabs.harvard.edu/abs/2011MNRAS.410.1550R} {410, 1550}

\bibitem[\protect\citeauthoryear{{Ramos Almeida} et~al.,}{{Ramos Almeida}
  et~al.}{2012}]{RamosAlmeida2012}
{Ramos Almeida} C.,  et~al., 2012, \mn@doi [\mnras]
  {10.1111/j.1365-2966.2011.19731.x}, \href
  {https://ui.adsabs.harvard.edu/abs/2012MNRAS.419..687R} {419, 687}

\bibitem[\protect\citeauthoryear{{Ramos Almeida} et~al.,}{{Ramos Almeida}
  et~al.}{2022}]{RamosAlmeida2022}
{Ramos Almeida} C.,  et~al., 2022, \mn@doi [\aap]
  {10.1051/0004-6361/202141906}, \href
  {https://ui.adsabs.harvard.edu/abs/2022A&A...658A.155R} {658, A155}

\bibitem[\protect\citeauthoryear{Rodriguez-Gomez et~al.,}{Rodriguez-Gomez
  et~al.}{2015}]{Rodriguez-Gomez2015}
Rodriguez-Gomez V.,  et~al., 2015, \mn@doi [\mnras] {10.1093/mnras/stv264},
  449, 49

\bibitem[\protect\citeauthoryear{{Roos} \& {Norman}}{{Roos} \&
  {Norman}}{1979}]{Roos1979}
{Roos} N.,  {Norman} C.~A.,  1979, \aap, \href
  {https://ui.adsabs.harvard.edu/abs/1979A&A....76...75R} {76, 75}

\bibitem[\protect\citeauthoryear{{Rosario} et~al.,}{{Rosario}
  et~al.}{2013a}]{Rosario20131}
{Rosario} D.~J.,  et~al., 2013a, \mn@doi [\aap] {10.1051/0004-6361/201322196},
  \href {https://ui.adsabs.harvard.edu/abs/2013A&A...560A..72R} {560, A72}

\bibitem[\protect\citeauthoryear{{Rosario}, {Burtscher}, {Davies}, {Genzel},
  {Lutz}  \& {Tacconi}}{{Rosario} et~al.}{2013b}]{Rosario20132}
{Rosario} D.~J.,  {Burtscher} L.,  {Davies} R.,  {Genzel} R.,  {Lutz} D.,
  {Tacconi} L.~J.,  2013b, \mn@doi [\apj] {10.1088/0004-637X/778/2/94}, \href
  {https://ui.adsabs.harvard.edu/abs/2013ApJ...778...94R} {778, 94}

\bibitem[\protect\citeauthoryear{{Rosario} et~al.,}{{Rosario}
  et~al.}{2015}]{Rosario2015}
{Rosario} D.~J.,  et~al., 2015, \mn@doi [\aap] {10.1051/0004-6361/201423782},
  \href {https://ui.adsabs.harvard.edu/abs/2015A&A...573A..85R} {573, A85}

\bibitem[\protect\citeauthoryear{{Rowan-Robinson}}{{Rowan-Robinson}}{1995}]{RowanRobinson1995}
{Rowan-Robinson} M.,  1995, \mn@doi [\mnras] {10.1093/mnras/272.4.737}, \href
  {https://ui.adsabs.harvard.edu/abs/1995MNRAS.272..737R} {272, 737}

\bibitem[\protect\citeauthoryear{{Rowlands}, {Wild}, {Nesvadba}, {Sibthorpe},
  {Mortier}, {Lehnert}  \& {da Cunha}}{{Rowlands} et~al.}{2015}]{Rowlands2015}
{Rowlands} K.,  {Wild} V.,  {Nesvadba} N.,  {Sibthorpe} B.,  {Mortier} A.,
  {Lehnert} M.,   {da Cunha} E.,  2015, \mn@doi [\mnras]
  {10.1093/mnras/stu2714}, \href
  {https://ui.adsabs.harvard.edu/abs/2015MNRAS.448..258R} {448, 258}

\bibitem[\protect\citeauthoryear{{Saito} et~al.,}{{Saito}
  et~al.}{2022}]{Saito2022}
{Saito} T.,  et~al., 2022, \mn@doi [\apjl] {10.3847/2041-8213/ac59ae}, \href
  {https://ui.adsabs.harvard.edu/abs/2022ApJ...927L..32S} {927, L32}

\bibitem[\protect\citeauthoryear{{Sanders}, {Soifer}, {Elias}, {Madore},
  {Matthews}, {Neugebauer}  \& {Scoville}}{{Sanders}
  et~al.}{1988}]{Sanders1988}
{Sanders} D.~B.,  {Soifer} B.~T.,  {Elias} J.~H.,  {Madore} B.~F.,  {Matthews}
  K.,  {Neugebauer} G.,   {Scoville} N.~Z.,  1988, \mn@doi [\apj]
  {10.1086/165983}, \href
  {https://ui.adsabs.harvard.edu/abs/1988ApJ...325...74S} {325, 74}

\bibitem[\protect\citeauthoryear{{Santini} et~al.,}{{Santini}
  et~al.}{2012}]{Santini2012}
{Santini} P.,  et~al., 2012, \mn@doi [\aap] {10.1051/0004-6361/201118266},
  \href {https://ui.adsabs.harvard.edu/abs/2012A&A...540A.109S} {540, A109}

\bibitem[\protect\citeauthoryear{{Satyapal}, {Ellison}, {McAlpine}, {Hickox},
  {Patton}  \& {Mendel}}{{Satyapal} et~al.}{2014}]{Satyapal2014}
{Satyapal} S.,  {Ellison} S.~L.,  {McAlpine} W.,  {Hickox} R.~C.,  {Patton}
  D.~R.,   {Mendel} J.~T.,  2014, \mn@doi [\mnras] {10.1093/mnras/stu650},
  \href {https://ui.adsabs.harvard.edu/abs/2014MNRAS.441.1297S} {441, 1297}

\bibitem[\protect\citeauthoryear{{Schawinski}, {Treister}, {Urry}, {Cardamone},
  {Simmons}  \& {Yi}}{{Schawinski} et~al.}{2011}]{Schawinski2011}
{Schawinski} K.,  {Treister} E.,  {Urry} C.~M.,  {Cardamone} C.~N.,  {Simmons}
  B.,   {Yi} S.~K.,  2011, \mn@doi [\apjl] {10.1088/2041-8205/727/2/L31}, \href
  {https://ui.adsabs.harvard.edu/abs/2011ApJ...727L..31S} {727, L31}

\bibitem[\protect\citeauthoryear{{Schawinski}, {Simmons}, {Urry}, {Treister}
  \& {Glikman}}{{Schawinski} et~al.}{2012}]{Schawinski2012}
{Schawinski} K.,  {Simmons} B.~D.,  {Urry} C.~M.,  {Treister} E.,   {Glikman}
  E.,  2012, \mn@doi [\mnras] {10.1111/j.1745-3933.2012.01302.x}, \href
  {https://ui.adsabs.harvard.edu/abs/2012MNRAS.425L..61S} {425, L61}

\bibitem[\protect\citeauthoryear{{Schmidt}}{{Schmidt}}{1959}]{Schmidt1959}
{Schmidt} M.,  1959, \mn@doi [\apj] {10.1086/146614}, \href
  {https://ui.adsabs.harvard.edu/abs/1959ApJ...129..243S} {129, 243}

\bibitem[\protect\citeauthoryear{{Schweitzer} et~al.,}{{Schweitzer}
  et~al.}{2006}]{Schweitzer2006}
{Schweitzer} M.,  et~al., 2006, \mn@doi [\apj] {10.1086/506510}, \href
  {https://ui.adsabs.harvard.edu/abs/2006ApJ...649...79S} {649, 79}

\bibitem[\protect\citeauthoryear{{Scudder}, {Ellison}, {Torrey}, {Patton}  \&
  {Mendel}}{{Scudder} et~al.}{2012}]{Scudder2012}
{Scudder} J.~M.,  {Ellison} S.~L.,  {Torrey} P.,  {Patton} D.~R.,   {Mendel}
  J.~T.,  2012, \mn@doi [\mnras] {10.1111/j.1365-2966.2012.21749.x}, \href
  {https://ui.adsabs.harvard.edu/abs/2012MNRAS.426..549S} {426, 549}

\bibitem[\protect\citeauthoryear{{Scudder}, {Ellison}, {Momjian}, {Rosenberg},
  {Torrey}, {Patton}, {Fertig}  \& {Mendel}}{{Scudder}
  et~al.}{2015}]{Scudder2015}
{Scudder} J.~M.,  {Ellison} S.~L.,  {Momjian} E.,  {Rosenberg} J.~L.,  {Torrey}
  P.,  {Patton} D.~R.,  {Fertig} D.,   {Mendel} J.~T.,  2015, \mn@doi [\mnras]
  {10.1093/mnras/stv588}, \href
  {https://ui.adsabs.harvard.edu/abs/2015MNRAS.449.3719S} {449, 3719}

\bibitem[\protect\citeauthoryear{{Secrest}, {Ellison}, {Satyapal}  \&
  {Blecha}}{{Secrest} et~al.}{2020}]{Secrest2020}
{Secrest} N.~J.,  {Ellison} S.~L.,  {Satyapal} S.,   {Blecha} L.,  2020,
  \mn@doi [\mnras] {10.1093/mnras/staa1692}, \href
  {https://ui.adsabs.harvard.edu/abs/2020MNRAS.499.2380S} {499, 2380}

\bibitem[\protect\citeauthoryear{{Shah} et~al.,}{{Shah}
  et~al.}{2020}]{Shah2020}
{Shah} E.~A.,  et~al., 2020, \mn@doi [\apj] {10.3847/1538-4357/abbf59}, \href
  {https://ui.adsabs.harvard.edu/abs/2020ApJ...904..107S} {904, 107}

\bibitem[\protect\citeauthoryear{{Shangguan}, {Ho}, {Bauer}, {Wang}  \&
  {Treister}}{{Shangguan} et~al.}{2020}]{Shangguan2020}
{Shangguan} J.,  {Ho} L.~C.,  {Bauer} F.~E.,  {Wang} R.,   {Treister} E.,
  2020, \mn@doi [\apj] {10.3847/1538-4357/aba8a1}, \href
  {https://ui.adsabs.harvard.edu/abs/2020ApJ...899..112S} {899, 112}

\bibitem[\protect\citeauthoryear{{Shao} et~al.,}{{Shao}
  et~al.}{2010}]{Shao2010}
{Shao} L.,  et~al., 2010, \mn@doi [\aap] {10.1051/0004-6361/201014606}, \href
  {https://ui.adsabs.harvard.edu/abs/2010A&A...518L..26S} {518, L26}

\bibitem[\protect\citeauthoryear{{Smercina} et~al.,}{{Smercina}
  et~al.}{2022}]{Smercina2022}
{Smercina} A.,  et~al., 2022, \mn@doi [\apj] {10.3847/1538-4357/ac5d5f}, \href
  {https://ui.adsabs.harvard.edu/abs/2022ApJ...929..154S} {929, 154}

\bibitem[\protect\citeauthoryear{{Springel}, {Di Matteo}  \&
  {Hernquist}}{{Springel} et~al.}{2005}]{Springel2005}
{Springel} V.,  {Di Matteo} T.,   {Hernquist} L.,  2005, \mn@doi [\mnras]
  {10.1111/j.1365-2966.2005.09238.x}, \href
  {https://ui.adsabs.harvard.edu/abs/2005MNRAS.361..776S} {361, 776}

\bibitem[\protect\citeauthoryear{Springel et~al.,}{Springel
  et~al.}{2017}]{Springel2017}
Springel V.,  et~al., 2017, \mn@doi [\mnras] {10.1093/mnras/stx3304}, 475, 676

\bibitem[\protect\citeauthoryear{{Steinborn}, {Hirschmann}, {Dolag}, {Shankar},
  {Juneau}, {Krumpe}, {Remus}  \& {Teklu}}{{Steinborn}
  et~al.}{2018}]{Steinborn2018}
{Steinborn} L.~K.,  {Hirschmann} M.,  {Dolag} K.,  {Shankar} F.,  {Juneau} S.,
  {Krumpe} M.,  {Remus} R.-S.,   {Teklu} A.~F.,  2018, \mn@doi [\mnras]
  {10.1093/mnras/sty2288}, \href
  {https://ui.adsabs.harvard.edu/abs/2018MNRAS.481..341S} {481, 341}

\bibitem[\protect\citeauthoryear{Terrazas et~al.,}{Terrazas
  et~al.}{2020}]{Terrazas2020}
Terrazas B.~A.,  et~al., 2020, \mn@doi [\mnras] {10.1093/mnras/staa374}, 493,
  1888

\bibitem[\protect\citeauthoryear{{Thorp}, {Ellison}, {Simard}, {S{\'a}nchez}
  \& {Antonio}}{{Thorp} et~al.}{2019}]{Thorp2019}
{Thorp} M.~D.,  {Ellison} S.~L.,  {Simard} L.,  {S{\'a}nchez} S.~F.,
  {Antonio} B.,  2019, \mn@doi [\mnras] {10.1093/mnrasl/sly185}, \href
  {https://ui.adsabs.harvard.edu/abs/2019MNRAS.482L..55T} {482, L55}

\bibitem[\protect\citeauthoryear{{Toomre} \& {Toomre}}{{Toomre} \&
  {Toomre}}{1972}]{Toomre1972}
{Toomre} A.,  {Toomre} J.,  1972, \mn@doi [\apj] {10.1086/151823}, \href
  {https://ui.adsabs.harvard.edu/abs/1972ApJ...178..623T} {178, 623}

\bibitem[\protect\citeauthoryear{{Treister}, {Schawinski}, {Urry}  \&
  {Simmons}}{{Treister} et~al.}{2012}]{Treister2012}
{Treister} E.,  {Schawinski} K.,  {Urry} C.~M.,   {Simmons} B.~D.,  2012,
  \mn@doi [\apjl] {10.1088/2041-8205/758/2/L39}, \href
  {https://ui.adsabs.harvard.edu/abs/2012ApJ...758L..39T} {758, L39}

\bibitem[\protect\citeauthoryear{{Urbano-Mayorgas} et~al.,}{{Urbano-Mayorgas}
  et~al.}{2019}]{UrbanoMayorgas2019}
{Urbano-Mayorgas} J.~J.,  et~al., 2019, \mn@doi [\mnras]
  {10.1093/mnras/sty2910}, \href
  {https://ui.adsabs.harvard.edu/abs/2019MNRAS.483.1829U} {483, 1829}

\bibitem[\protect\citeauthoryear{{Villforth} et~al.,}{{Villforth}
  et~al.}{2014}]{Villforth2014}
{Villforth} C.,  et~al., 2014, \mn@doi [\mnras] {10.1093/mnras/stu173}, \href
  {https://ui.adsabs.harvard.edu/abs/2014MNRAS.439.3342V} {439, 3342}

\bibitem[\protect\citeauthoryear{{Villforth} et~al.,}{{Villforth}
  et~al.}{2017}]{Villforth2017}
{Villforth} C.,  et~al., 2017, \mn@doi [\mnras] {10.1093/mnras/stw3037}, \href
  {https://ui.adsabs.harvard.edu/abs/2017MNRAS.466..812V} {466, 812}

\bibitem[\protect\citeauthoryear{{Villumsen}}{{Villumsen}}{1982}]{Villumsen1982}
{Villumsen} J.~V.,  1982, \mn@doi [\mnras] {10.1093/mnras/199.3.493}, \href
  {https://ui.adsabs.harvard.edu/abs/1982MNRAS.199..493V} {199, 493}

\bibitem[\protect\citeauthoryear{{Violino}, {Ellison}, {Sargent}, {Coppin},
  {Scudder}, {Mendel}  \& {Saintonge}}{{Violino} et~al.}{2018}]{Violino2018}
{Violino} G.,  {Ellison} S.~L.,  {Sargent} M.,  {Coppin} K. E.~K.,  {Scudder}
  J.~M.,  {Mendel} T.~J.,   {Saintonge} A.,  2018, \mn@doi [\mnras]
  {10.1093/mnras/sty345}, \href
  {https://ui.adsabs.harvard.edu/abs/2018MNRAS.476.2591V} {476, 2591}

\bibitem[\protect\citeauthoryear{{Volonteri}, {Capelo}, {Netzer}, {Bellovary},
  {Dotti}  \& {Governato}}{{Volonteri} et~al.}{2015a}]{Volonteri20151}
{Volonteri} M.,  {Capelo} P.~R.,  {Netzer} H.,  {Bellovary} J.,  {Dotti} M.,
  {Governato} F.,  2015a, \mn@doi [\mnras] {10.1093/mnras/stv387}, \href
  {https://ui.adsabs.harvard.edu/abs/2015MNRAS.449.1470V} {449, 1470}

\bibitem[\protect\citeauthoryear{{Volonteri}, {Capelo}, {Netzer}, {Bellovary},
  {Dotti}  \& {Governato}}{{Volonteri} et~al.}{2015b}]{Volonteri20152}
{Volonteri} M.,  {Capelo} P.~R.,  {Netzer} H.,  {Bellovary} J.,  {Dotti} M.,
  {Governato} F.,  2015b, \mn@doi [\mnras] {10.1093/mnrasl/slv078}, \href
  {https://ui.adsabs.harvard.edu/abs/2015MNRAS.452L...6V} {452, L6}

\bibitem[\protect\citeauthoryear{{Ward}, {Harrison}, {Costa}  \&
  {Mainieri}}{{Ward} et~al.}{2022}]{Ward2022}
{Ward} S.~R.,  {Harrison} C.~M.,  {Costa} T.,   {Mainieri} V.,  2022, \mn@doi
  [\mnras] {10.1093/mnras/stac1219}, \href
  {https://ui.adsabs.harvard.edu/abs/2022MNRAS.514.2936W} {514, 2936}

\bibitem[\protect\citeauthoryear{{Weinberger} et~al.,}{{Weinberger}
  et~al.}{2017}]{Weinberger2017}
{Weinberger} R.,  et~al., 2017, \mn@doi [\mnras] {10.1093/mnras/stw2944}, \href
  {https://ui.adsabs.harvard.edu/abs/2017MNRAS.465.3291W} {465, 3291}

\bibitem[\protect\citeauthoryear{{Weston}, {McIntosh}, {Brodwin}, {Mann},
  {Cooper}, {McConnell}  \& {Nielsen}}{{Weston} et~al.}{2017}]{Weston2017}
{Weston} M.~E.,  {McIntosh} D.~H.,  {Brodwin} M.,  {Mann} J.,  {Cooper} A.,
  {McConnell} A.,   {Nielsen} J.~L.,  2017, \mn@doi [\mnras]
  {10.1093/mnras/stw2620}, \href
  {https://ui.adsabs.harvard.edu/abs/2017MNRAS.464.3882W} {464, 3882}

\bibitem[\protect\citeauthoryear{White}{White}{1978}]{White1978}
White S. D.~M.,  1978, \mn@doi [\mnras] {10.1093/mnras/184.2.185}, 184, 185

\bibitem[\protect\citeauthoryear{{Wild}, {Heckman}  \& {Charlot}}{{Wild}
  et~al.}{2010}]{Wild2010}
{Wild} V.,  {Heckman} T.,   {Charlot} S.,  2010, \mn@doi [\mnras]
  {10.1111/j.1365-2966.2010.16536.x}, \href
  {https://ui.adsabs.harvard.edu/abs/2010MNRAS.405..933W} {405, 933}

\bibitem[\protect\citeauthoryear{{Woods} \& {Geller}}{{Woods} \&
  {Geller}}{2007}]{Woods2007}
{Woods} D.~F.,  {Geller} M.~J.,  2007, \mn@doi [\aj] {10.1086/519381}, \href
  {https://ui.adsabs.harvard.edu/abs/2007AJ....134..527W} {134, 527}

\bibitem[\protect\citeauthoryear{{Woods}, {Geller}  \& {Barton}}{{Woods}
  et~al.}{2006}]{Woods2006}
{Woods} D.~F.,  {Geller} M.~J.,   {Barton} E.~J.,  2006, \mn@doi [\aj]
  {10.1086/504834}, \href
  {https://ui.adsabs.harvard.edu/abs/2006AJ....132..197W} {132, 197}

\bibitem[\protect\citeauthoryear{{Woods}, {Geller}, {Kurtz}, {Westra},
  {Fabricant}  \& {Dell'Antonio}}{{Woods} et~al.}{2010}]{Woods2010}
{Woods} D.~F.,  {Geller} M.~J.,  {Kurtz} M.~J.,  {Westra} E.,  {Fabricant}
  D.~G.,   {Dell'Antonio} I.,  2010, \mn@doi [\aj]
  {10.1088/0004-6256/139/5/1857}, \href
  {https://ui.adsabs.harvard.edu/abs/2010AJ....139.1857W} {139, 1857}

\bibitem[\protect\citeauthoryear{{Xie}, {Ho}, {Zhuang}  \& {Shangguan}}{{Xie}
  et~al.}{2021}]{Xie2021}
{Xie} Y.,  {Ho} L.~C.,  {Zhuang} M.-Y.,   {Shangguan} J.,  2021, \mn@doi [\apj]
  {10.3847/1538-4357/abe404}, \href
  {https://ui.adsabs.harvard.edu/abs/2021ApJ...910..124X} {910, 124}

\makeatother
\end{thebibliography}







\bsp	
\label{lastpage}
\end{document}